\documentclass[a4paper]{article}
\usepackage[left=1.5in, right=1.5in]{geometry}
\usepackage{authblk}

\usepackage[parfill]{parskip}

\usepackage[utf8x]{inputenc}

\usepackage{amsmath}
\usepackage{amssymb,amsmath,amsthm, mathtools,extarrows}
\usepackage{physics}
\usepackage{graphicx}
\usepackage[dvipsnames]{xcolor}
\usepackage{hyperref}
\hypersetup{breaklinks=true,
            pdfauthor={Michalis Michaelides},
            pdftitle={Geometric fluid approximation for general continuous-time Markov chains},
            colorlinks=true,
            citecolor=blue,
            urlcolor=blue,
            linkcolor=magenta
            }

\usepackage{subcaption}
\usepackage{tabularx}
\usepackage{algorithm}
\usepackage{algpseudocode}
\usepackage{bm}

\newtheorem{definition}{\bf Definition}[section]
\newtheorem{theorem}{\bf Theorem}[section]

\DeclareMathOperator*{\argmin}{argmin} 

\begin{document}

\title{Geometric fluid approximation for general continuous-time Markov chains}






\author[1]{Michalis Michaelides}
\author[1]{Jane Hillston}
\author[1, 2]{Guido Sanguinetti}

\affil[1]{\small School of Informatics, University of Edinburgh}
\affil[2]{\small SynthSys, Centre for Synthetic and Systems Biology, University of Edinburgh}




\maketitle

\begin{abstract}
Fluid approximations have seen great success in approximating the macro-scale behaviour of Markov systems with a large number of discrete states. However, these methods rely on the continuous-time Markov chain (CTMC) having a particular population structure which suggests a natural continuous state-space endowed with a dynamics for the approximating process.

We construct here a general method based on spectral analysis of the transition matrix of the CTMC, without the need for a population structure. Specifically, we use the popular manifold learning method of diffusion maps to analyse the transition matrix as the operator of a hidden continuous process. An embedding of states in a continuous space is recovered, and the space is endowed with a drift vector field inferred via Gaussian process regression. In this manner, we construct an ODE whose solution approximates the evolution of the CTMC mean, mapped onto the continuous space (known as the fluid limit).
\end{abstract}


\section{Introduction}
Stochastic process models of dynamical systems play a central role in scientific investigations across a broad range of disciplines, from computer science, to physics, to biology. Continuous-time Markov chains (CTMCs), in particular, have emerged over the last two decades as an especially powerful class of models to capture the intrinsic discreteness and stochasticity of biological systems at the single cell level. The ensuing cross-fertilisation of stochastic systems biology with methods emerging from the formal modelling of computer systems has led to the dramatic explosion of a new interdisciplinary field at the intersection of computer science and biology \cite{schnoerr_approximation_2017,hillston_fluid_2005,larsson_genomic_2019}.

Despite the unquestionable success of these modelling efforts, scaling formal analysis techniques to larger systems remains a major challenge, since such systems usually result in very large state-spaces, making subsequent analysis particularly onerous.
In most cases, retrieving the evolution of the state distribution, while theoretically possible (by solving the Kolmogorov-Chapman equations),  in practice is prohibitively expensive. These hurdles also affect statistical techniques based on Monte Carlo sampling, since trajectories from CTMCs with a large state-space typically exhibit very frequent transitions and therefore require the generation of a very large number of random numbers. A popular alternative is therefore to rely on {\it model approximations}, by constructing alternative models which in some sense approximate the system's behaviour. In the special case of CTMCs with a population structure (pCTMCs), \emph{fluid approximations} replacing the original dynamics with a deterministic set of ordinary differential equations (ODEs) have seen great success, due to their scalability and their well understood convergence properties \cite{kurtz_limit_1971,hillston_fluid_2005,darling_differential_2008}. Such approximations rely on the particular structure of the state-space of pCTMCs; to the best of our knowledge, fluid approximations for general CTMCs have not been developed.

In this paper, we propose a general strategy to obtain a fluid approximation for any CTMC. Our approach utilises manifold learning approaches, popular in machine learning, to embed the transition graph of a general CTMC in a Euclidean space. A powerful Bayesian non-parametric regression method complements the embedding, by inferring a drift vector field over the Euclidean space to yield a continuous process, which we term the \emph{geometric fluid approximation} (GFA). The accuracy of the GFA depends on how naturally the CTMC embeds in a Euclidean space. Crucially, we show that in a simple pCTMC case an approximation related to the GFA is consistent with the standard fluid approximation. Empirical results on a range of examples of CTMCs without a strict population structure show that if the transition graph approximately resembles a continuous manifold, our approach captures well the average behaviour of the CTMC trajectories, and can be useful to efficiently solve approximate reachability problems.

\section{Background theory and related work}
We briefly review here the mathematical definition of CTMCs, as well as the foundations of fluid approximation for population CTMCs (pCTMCs) \cite{gardiner_stochastic_2009,norris_markov_1998,darling_differential_2008}, highlighting the specific aspects of pCTMCs that enable such a construction.

\subsection{Continuous time Markov chains}
A {\it continuous time Markov Chain (CTMC)} is a continuous-time, Markovian stochastic process over a finite state-space $I$. The process is characterised by its {\it generator matrix} $Q$, encoding the infinitesimal probability of transitioning between any two states. A more formal definition of CTMC
 \cite{norris_markov_1998} can be given as follows:
\begin{definition}\label{def:ctmc}
Let $X(t),~ t\in T$, be a right-continuous process with values in a countable set $I\subset \mathbb{N}_{>0}$. Let $Q$ be a generator matrix on $I$ with jump matrix $\Pi$, such that for $i, j \in I$:
\begin{align*}
&Q_{ij}\in \mathbb{R}_{\geq 0}   \qquad \forall i \neq j, \qquad Q_{ii} = - \sum_{j\neq i} Q_{ij}   \qquad \forall i, \\
\text{and} \qquad
&\Pi_{ij} = \begin{cases}
            - Q_{ij}/ Q_{ii} &\quad\text{if } i \neq j ~ \land~ Q_{ii} \neq 0, \\
            0 &\quad\text{otherwise}.
           \end{cases}
\end{align*}
The process $X(t)$ is a \emph{continuous-time Markov chain} with initial state distribution $\pi$ and generator matrix $Q$ (CTMC($\pi, Q$)), if it satisfies:
\begin{enumerate}
\item{}$P(X(0) = i) = \pi_i$; and
\item{}$X(t) = Y_n$ for $\sum_{k=0}^{n} S_k \geq t \geq \sum_{k=0}^{n-1} S_k$, where $n\in \mathbb{N}_{>0}$ and $S_0 = 0$,\\
such that $P(Y_{n+1}=j | Y_{n}=i) = \Pi_{ij}$, and $\forall~n \geq 1, S_n \mid Y_{n} = i \sim \text{Exp}(-Q_{ii})$.
\end{enumerate}
\end{definition}
This definition emphasises the so-called \emph{holding (or residence) times}, i.e.\ the random time that the system spends into any one state. The exponential distribution of the holding times is a simple consequence of the Markovian nature of the process; it also naturally suggests an exact algorithm to sample trajectories of CTMCs by drawing repeatedly exponential random numbers. This consideration forms the basis of the \emph{Gillespie algorithm}, widely used in the field of systems biology and known as the \emph{stochastic simulation algorithm} (SSA) \cite{gillespie_exact_1977}.

An important special instance of the CTMC is the so-called population CTMC (pCTMC). Population CTMCs model systems consisting of indistinguishable agents of different types, or \emph{species}. The state-space is therefore identified with a vector of integer numbers, with each entry counting the species population, and transitions occurring at different rates depending on the counts of each agent type.

The transitions in a pCTMC of $m \in \mathbb{N}_{>0}$ species can be regarded as occurrences of $R$ chemical reactions, written as
\[
\sum_{i=1}^m u_{ir} \hat{\mathbf{e}}_i \xrightarrow{k_r} \sum_{i=1}^m v_{ir} \hat{\mathbf{e}}_i, \qquad r = 1,\dots, R,
\]
where $u_{ir}$ counts the number of particles of species $\hat{\mathbf{e}}_i$ that are consumed and $v_{ir}$ the particles of the same species created, in reaction $r$ (species unit vectors are orthonormal: $\langle\hat{\mathbf{e}}_i | \hat{\mathbf{e}}_j\rangle = \delta_{i, j}$). The \emph{reaction rate constant} $k_r$ is a factor in the \emph{propensity function} $f_r(\mathbf{u})$, which constructs the $Q$-matrix of the CTMC. See \cite{gardiner_stochastic_2009, schnoerr_approximation_2017} for a thorough exposition of pCTMCs for chemical reaction networks and a justified definition of the propensity function.

\subsection{Continuous relaxation and the \emph{fluid limit}}
As discussed in the previous section, the Markovian nature of CTMCs naturally provides an exact sampling algorithm for drawing CTMC trajectories. The same Markovian nature also leads to a set of ordinary differential equations (ODEs) governing the evolution of the single-time marginal state probability, the celebrated \emph{Chapman-Kolmogorov equations} (CKE), which in the case of pCTMCs go under the name of \emph{Master equation}. Unfortunately, such equations are rarely practically solvable, and analysis of CTMCs is often reliant on computationally intensive simulations.

In the case of pCTMCs, a more concise description in terms of the collective dynamics of population averages is however available. Starting with the seminal work of van Kampen \cite{kampen_power_1961}, and motivated by the interpretation of pCTMCs as chemical reaction systems, several approximation schemes have been developed which relax the original pCTMC to a continuous stochastic process; see \cite{schnoerr_approximation_2017} for a recent review.

In this paper, we are primarily interested in the so-called \emph{fluid approximation}, which replaces the pCTMC with a set of ODEs, which captures the average behaviour of the system. Fluid approximations have been intensely studied and their limiting behaviour is well understood, providing specific error guarantees.
There are two characteristics of a pCTMC which are instrumental to enabling the fluid approximation. Firstly, there is a natural interpretation of the states as points in a vector space, where each dimension represents a species. Secondly, a \emph{drift vector field} can be naturally defined by extending the propensity function to be defined on the whole vector space, which is a polynomial function of the number of agents of each type (i.e.\ polynomial function of the elements of the system state vector).

\paragraph{Established guarantees}\label{gfa:sec:theory:established}
Following Darling and Norris \cite{darling_differential_2008}, we examine and formalise the aspects of pCTMCs which render them especially amenable to the fluid approximation. As mentioned, the first is that pCTMC state-spaces are countable and there exists an obvious ordering. We can therefore write a trivial linear mapping from the discrete, countable state-space $I$ to a continuous Euclidean space $\mathbf{x}:I \to \mathbb{R}^d$, where $d$ is the number of agent types in the system.

The second aspect is that rates of transition from each state to all others (i.e.\ elements of the $Q$-matrix) can be expressed as a function of the state vector $\mathbf{x}$. A \emph{drift vector} $\beta(\xi)$ can be defined as
\[
\beta(\xi) = \sum_{\xi'\neq\xi} \left(\mathbf{x}(\xi') - \mathbf{x}(\xi)\right)q(\xi, \xi'),
\]
for each $\xi \in I$. Since $q(\xi, \xi')$ is some parametric function of $\xi, \xi'$ in pCTMCs (due to the indistinguishable nature of the agents) the definition of the \emph{drift vector} can be extended over the entire Euclidean space $\mathbb{R}^d$ to produce the \emph{drift vector field} $b(\mathbf{x}): U \to \mathbb{R}^d$, where $U\subseteq \mathbb{R}^d$. There is then a set of conditions given in \cite{darling_differential_2008} that must be satisfied by these elements to bind the error of the fluid approximation to the Markov process. The conditions ensure that:

\emph{The first exit time from a suitably selected domain of the Euclidean mapping of the Markov chain state-space $U$, converges in probability to the first exit time of the fluid limit.}

\paragraph{Canonical embedding of pCTMCs}
In the canonical embedding for continuous relaxation of pCTMCs, we construct an $E \subset \mathbb{R}^d$ Euclidean space, where each dimension corresponds to the concentration of each species in the system, $i\in\{1, \dots, m\}$. The states are then uniformly embedded in continuous space $[0,1]^m \in E$ at intervals $1/n_i$ by $\mathbf{x}(\xi) = u_i / n_i$, where $\xi$ represents the population $\sum_i u_i \hat{\mathbf{e}}_i$. Further, $N = |I| = \prod_i n_i$, is a scale parameter which defines $\mathbf{x}_N(\xi)$, $q_N(\xi, \xi')$ and $\beta_N(\xi)$ for any such pCTMC of size $N$. The motivation is that in the limit of $N \to \infty$, the distance between neighbouring states will vanish in the embedding, and jump sizes will similarly vanish, producing an approximately continuous trajectory of the system in the continuous concentration space.

In \cite{darling_fluid_2002}, we find how the canonical embedding above satisfies the conditions given in \cite{darling_differential_2008}, and that the approximation error shrinks as the scale parameter $N$ grows. Specifically, the authors show that there exists a fluid approximation (deterministic trajectory) to the $\mathbf{x}$-mapped pCTMC, whose error diminishes in $N$, under the conditions that:
\begin{itemize}
    \item{\emph{initial conditions} converge}, i.e.\ $\exists a \in U,\ a \neq \mathbf{x}(\xi_0)$ such that
    \[ \Pr\left[ \lVert \mathbf{x}_N(\xi_0) - a \rVert > \delta\right] \leq \kappa_1(\delta) / N, \quad
    \forall \delta>0;\]

    \item{\emph{mean dynamics} converge as $N\to\infty$,} i.e.\
    $\tilde{b}:U \to \mathbb{R}^d$ is a Lipschitz field independent of $N$, such that
    \[\sup_{\xi} \lVert \beta_N(\xi) - \tilde{b}(\mathbf{x}_N(\xi)) \rVert \to 0 \quad \text{as}\  N\to\infty;
    \]

    \item{\emph{noise} converges to zero as $N\to\infty$,} i.e.\ that,
    \begin{align*}
    &\sup_{\xi} \left\{ \sum_{\xi'\neq\xi}q_N(\xi, \xi') \right\} \leq \kappa_2 N,
    	\quad \text{and}\\
    &\sup_{\xi} \left\{
    \lVert\beta_N(\xi)/q_N(\xi)\rVert^2 +
    \sum_{\xi'\neq\xi} \lVert \mathbf{x}_N(\xi') - \mathbf{x}_N(\xi) \rVert^2 q_N(\xi, \xi') / q_N(\xi) \right\} \leq \kappa_3 N^{-2},
    \end{align*}
\end{itemize}
where $\kappa_1(\delta),~\kappa_2,~\kappa_3$ are positive constants, $q(\xi) = \sum_{\xi'\neq \xi} q(\xi, \xi')$, and the inequalities hold uniformly in $N$.

There are many ways to satisfy the above criteria, but a common one (used in pCTMCs) is ``hydrodynamic scaling'', where the increments of the $N$-state Markov process mapped to the Euclidean space are $\mathcal{O}(N^{-1})$ and the jump rate is $\mathcal{O}(N)$.

\section{Methodology}\label{sec:method}

As discussed in the previous section, the fluid approximation of pCTMCs is critically reliant on the structure of the state-space of pCTMCs being isomorphic to a lattice in $\mathbb{R}^n$. This enables the definition of a drift vector field, which can be then naturally extended to the whole ambient space and, under mild assumptions, leads to convergence under suitable scaling. Neither of these ingredients are obviously available in the general case of CTMCs lacking a population structure.

In this section, we describe the proposed methodology for a geometric fluid approximation for CTMCs. We motivate our approach by describing an exact, if trivial, general embedding of a CTMC's state-space into a very high-dimensional space. Such an embedding however affords the non-trivial insight that suitable approximate embeddings may be obtained considering the spectral geometry of the generator matrix. This provides an unexpected link with a set of techniques from machine learning, {\it diffusion maps}, which embed graphs into Euclidean spaces. The geometry of diffusion maps is well studied, and their distance preservation property is particularly useful for our purpose of obtaining a fluid approximation.

Diffusion maps however provide only one ingredient to a fluid approximation; they do not define an ODE flow over the ambient Euclidean space. To do so, we use Gaussian Process regression: this provides a smooth interpolation of the dynamic field between embedded states. Smoothness guarantees that nearby states in the CTMC (which are embedded to nearby points in Euclidean space by virtue of the distance preservation property of diffusion maps) will have nearby drift values, somewhat enforcing the pCTMC property that the transition rates are a function of the state vector.

This two-step strategy provides a general approach to associate a deterministic flow on a vector space to a CTMC. We empirically validate that such flow indeed approximates the mean behaviour of the CTMCs on a range of examples in the next section. Prior to the empirical section, we prove a theorem showing that, in the special case of pCTMCs of birth/death type, our geometric construction returns the same fluid approximation as the standard pCTMC construction, providing an important consistency result.

{
\subsection{Eigen-embeddings of CTMCs}
\paragraph{Trivial embedding of CTMCs}
Consider a CTMC with initial distribution $\pi$ and generator matrix $Q$, on countable state-space $\Xi \subset \mathbb{N}$. The single time marginal $p_t$ over $\Xi$ at time $t$ of the process obeys the CKE:
\begin{equation}
\partial_t p_t = Q^\top p_t,
\end{equation}
where $p_t$ is a column vector. Given an arbitrary embedding of the states in some continuous space, $x: \Xi \to \mathbb{R}^d$, the projected mean $\langle x_t \rangle = X^\top p_t$, obeys:
\[
\partial_t \langle x_t \rangle = X^\top \partial_t p_t = X^\top Q^\top p_t
    = X^\top Q^\top X^{-\top} X^\top p_t
    = X^\top Q^\top X^{-\top} \langle x_t \rangle,
\]
where $X_{ij}$ refers to the $j\in \{1,\dots, d\}$ coordinate of state $i \in \Xi$. In general, the last step is only possible for $XX^{-1} = I$, with $I$ the $|\Xi| \times |\Xi|$ identity matrix (i.e.\ with $d = |\Xi|$).

We note that choosing the trivial embedding $X = I$ (i.e.\ each state mapped to the vertex of the probability ($|\Xi| - 1$)-simplex), equates the fluid process to the original CKE:
\begin{equation}
\partial_t \langle x_t \rangle = Q^\top \langle x_t \rangle.
\end{equation}

\paragraph{The fluid approximation}

For any embedding $y: \Xi \to \mathbb{R}^d$, the standard fluid approximation defines the drift at any state $y(i) \equiv y_i,\ i \in \Xi$, to be:
\[
\beta(y_i) = \sum_{j\neq i} (y_j - y_i) Q_{ij}
    = \sum_{j\neq i} y_j Q_{ij} - y_i \sum_{j\neq i} Q_{ij}= \sum_{j\neq i} y_j Q_{ij} + y_i Q_{ii} = [Q Y]_i,
\]
where $Y$ is a $|\Xi| \times d$ matrix, and $\beta(y_i)$ is the $i$th row of $Q Y$.

In order to extend the drift over the entire space $\mathbb{R}^d$, we let $q(y_t, y_j) = \sum_i Q_{ij} y_i^\top y_t$ be the transition kernel between any point $y_t \in \mathbb{R}^d$ and any state $y_j,\ j \in \Xi$. Then we naturally define the continuous vector field $b$ to be
\begin{align}
b(y_t) &= \sum_j y_j q(y_t, y_j)
    = \sum_j y_j \sum_i Q_{ij} y_i^\top y_t
    = Y^\top Q^\top Y y_t.
\end{align}

Trivially, $Y = I$ yields the original CKE as shown above,
\begin{equation}
\partial_t y_t = b(y_t) = Y^\top Q^\top Y y_t = Q^\top y_t.
\end{equation}

If embedded states $V$ are eigenvectors of $Q = V \Lambda V^{-1}$, then the mean of the mapped process $\langle v_t \rangle$ is given by
\begin{align}
\langle v_t \rangle &= V^\top p_t
    = V^\top V^{-\top} e^{t \Lambda} V^\top \pi
    = e^{t \Lambda} V^\top \pi.
\end{align}

Similarly, in the fluid approximation with $q(y_t, y_j) = \sum_i Q_{ij} y_i^\top y_t$, and $y_0 = V^\top \pi$, one has
\begin{align}
y_t = e^{t \Lambda V^\top V} y_0 = e^{t \Lambda V^\top V} V^\top \pi.
\end{align}
We can therefore claim the following: \emph{in the case of $V^\top V = I$, or for a symmetric $Q = V\Lambda V^\top$, the fluid approximation process is exactly equivalent to the projected mean.} This suggests that an approximate, low dimensional representation might be obtained by truncating the spectral expansion of the generator matrix of the CTMC.  Spectral analysis of a transport operator is also the approach taken by \emph{diffusion maps}, a method which is part of the burgeoning field of manifold learning for finding low-dimensional representations of high-dimensional data.

\paragraph{Markov chain as a random walk on a graph} Another avenue to reach the same conclusion is to consider the CTMC as a random walk on an abstract graph, where vertices represent states of the chain, and weighted directed edges represent possible transitions. From this perspective, it is natural to seek an embedding of the graph in a suitable low-dimensional vector space; it is well known in the machine learning community that an optimal (in a sense specified in Section~\ref{sec:method:diff}) embedding can be obtained by spectral analysis of the transport operator linked to the random walk on the graph. Intuitively, we expect that if the graph geometry discretely approximates some continuous space, then the underlying space will serve well as a continuous state-space for a fluid limit approximation, when endowed with an appropriate drift vector field to capture the non-geometric dynamics in the graph.

}

\subsection{Diffusion maps}\label{sec:method:diff}
A natural method to embed the CTMC states in continuous space for our purposes is \emph{diffusion maps} \cite{coifman_geometric_2005,coifman_diffusion_2006,nadler_diffusion_FP_2006,nadler_diffusion_react_2006,coifman_diffusion_stoch_2008}. This is a manifold learning method, where the authors consider a network defined by a symmetric adjacency matrix, with the aim of finding coordinates for the network vertices on a continuous manifold (as is usually the case with similarities of high-dimensional points).

\paragraph{Diffusion on a manifold}
The method of \emph{diffusion maps} follows from regarding high-dimensional points in $\mathbb{R}^p$ to be observations of a diffusion process at regular intervals, which evolves on a hidden manifold $\mathcal{M} \subset \mathbb{R}^p$ with a smooth boundary $\partial \mathcal{M}$. Concurrently, a similarity matrix between the high-dimensional points (i.e.\ an adjacency matrix) is interpreted as the un-normalised transition kernel of the hidden diffusion process. These assumptions imply that the geometry of $\mathcal{M}$ must be such that \emph{similar} points are likely consecutive observations of the diffusion process, since the latter is dependent on the geometry of $\mathcal{M}$. The goal is then to recover coordinates for the points, natural to their position on the assumed manifold --- in doing so, we infer a continuum $\mathcal{M}$ from relations between some of its point elements. In essence, we seek a low-dimensional representation of the points which best preserves their similarities as proximity.

In the context of CTMCs the points that are to be embedded are the CTMC states, and the transition matrix $Q$ is the transition kernel of the diffusion process evaluated at discrete points (CTMC states). A family of diffusion operators are constructed which can be spectrally analysed to yield coordinates for each vertex on the manifold. The continuous operators, which are theoretically constructed to govern the diffusion process, are assumed to be approximated by the analogous discrete operators which are constructed from data. The method can be thought to optimally preserve the normalised \emph{diffusion distance} of the diffusion process on the high-dimensional manifold, as Euclidean distance in the embedding. \emph{Diffusion distance} between two vertices $\mathbf{x}_0, \mathbf{x}_1$ at time $t$ is defined to be the distance between the probability densities over the state-space, each initialised at $\mathbf{x}_0, \mathbf{x}_1$ respectively, and after a time $t$ has passed:
\[
D_t^2 (\mathbf{x}_0, \mathbf{x}_1) = \lVert p(\mathbf{x}, t|\mathbf{x}_0) - p(\mathbf{x}, t|\mathbf{x}_1) \rVert^2_{L_2(w)},
\]
where $L_2(\mathcal{M}, w)$ is a Hilbert space in which the distance is defined, with $w(\mathbf{x}) = 1 / \phi_0(\mathbf{x})$, the inverse of the steady-state distribution $\phi_0(\mathbf{x}) = \lim_{t\to\infty} p(\mathbf{x}, t|\mathbf{x}_0, 0)$. The procedure is similar to principal component analysis, since taking the first $k<p$ eigenvectors of the diffusion distance matrix provides coordinates for nodes in a low dimensional space optimally preserving $\sum_{i, j} D_t^2 (\mathbf{x}_i, \mathbf{x}_j)$. We point to \cite{coifman_geometric_2005,coifman_diffusion_2006} for a comprehensive theoretical exposition to \emph{diffusion maps}.

\paragraph{Diffusion with drift for asymmetric networks}
The methodology of diffusion maps has been extended in \cite{perrault-joncas_directed_2011} to deal with learning manifold embeddings for directed weighted networks. Given an asymmetric adjacency matrix, the symmetric part is extracted and serves as a discrete approximation to a geometric operator on the manifold. Spectral analysis of the relevant matrix can then yield embedding coordinates for the nodes of the network. In the same manner as for the original formulation of diffusion maps a set of backward evolution operators are derived, the two relevant ones being:
\begin{align}
&-\partial_t \psi_t = \mathcal{H}_{aa}^{(\alpha)} \psi_t = \left[\Delta + \left(\mathbf{r} - 2(1 - \alpha)\nabla U \right)\cdot \nabla\right] \psi_t, \\
\text{and} \qquad &-\partial_t \psi_t = \mathcal{H}_{ss}^{(\alpha)} \psi_t = \left[\Delta - 2(1 - \alpha)\nabla U \cdot \nabla\right] \psi_t,
\end{align}
where $\psi_t \in \mathcal{C}^2(\mathcal{M})$ is the mean of a real-valued bounded function of a random walker on the manifold after time $t$ (e.g.\ $\psi_t$ is a probability density). The sampling potential $U$ defines the steady-state distribution of the diffusion, $\lim_{t\to\infty}p_t = e^{-U}/Z$ which is taken to represent the sampling density of points on the manifold. The operators are parametrised by $\alpha$, which determines how affected the diffusion process on the manifold is by $U$. Choosing $\alpha=1$ allows us to spectrally analyse a discrete approximation to the Laplace-Beltrami operator $\Delta = \mathcal{H}_{ss}^{(\alpha=1)}$, extricating the geometry of the manifold from the density dependent term $-2(1-\alpha)\nabla U \cdot \nabla$ in the diffusion operator $\mathcal{H}_{ss}^{(\alpha)}$. The choice of $\alpha$ effectively allows one to control how much the recovered Euclidean representation of the manifold geometry is affected by the sampling density. Finally, $\mathbf{r}$ is a drift vector component tangential to the manifold, which additively guides the diffusion process. Perrault-Joncas and Meil\u{a} comprehensively treat the application diffusion maps on directed graphs in \cite{perrault-joncas_directed_2011}.

\paragraph{Diffusion maps for CTMCs}
For an arbitrary CTMC($\pi, Q$), we regard $Q \in \mathbb{R}^{N\times N}$ to be a discrete approximation of the operator $\mathcal{H}_{aa}^{(\alpha)}$. However, it is unclear how one can extract the geometrically relevant component $\Delta$ under a hidden potential $U$ and parameter $\alpha$. In practice, therefore, we assume a uniform measure on the manifold, i.e.\ constant $U$, which renders $Q$ a discrete approximation of $\mathcal{H}_{aa} = \Delta + \mathbf{r} \cdot \nabla$ (the choice of $\alpha$ no longer matters); further, we take the sampling transition kernel corresponding to this operator to be composed of a symmetric and anti-symmetric part (without loss of generality), which renders $\lim_{N\to\infty}(Q + Q^T) / 2 = \tilde{\Delta}$, an un-normalised version of $\Delta = \text{diag}(\beta_1)\ \tilde{\Delta}\ \text{diag}(\beta_2)$.\footnote{Recall that $\mathbf{r} \cdot \nabla \psi$ is the drift vector component tangential to the manifold. As such, it is an anti-symmetric field in the limit $N\to\infty$, and so $\mathbf{r} \cdot \nabla$ must be an anti-symmetric operator under transposition when $N<\infty$.} $\Delta$ contains the relevant geometric information about the network, with the first $k + 1$ eigenvectors of the operator used as embedding coordinates in a $k$-dimensional Euclidean space (ignoring the first eigenvector which is trivial by construction). A detailed exposition of the method as it relates to our purposes of embedding a Markov chain network can be found in Appendix~\ref{app:dm-directed}.

It should be noted that, while diffusion maps have been used to construct low-dimensional approximations of high-dimensional SDEs \cite{coifman_diffusion_stoch_2008}, and to embed a discrete-time Markov chain in continuous space with an accompanying \emph{advective field} \cite{perrault-joncas_directed_2011}, doing the same for a continuous-time Markov chain has not been attempted. Distinctively, the focus of that work was not to clear a path between discrete and continuous state Markov processes, but rather the low-dimensional embedding of processes or sample points.
In terms of the convenient table presented in \cite{nadler_diffusion_FP_2006} and restated here in Table~\ref{tab:limits}, we seek to examine the omitted entry that completes the set of Markov models; this is the third entry added here to the original table, taking $N<\infty$ and the time interval between transitions limit $\epsilon\to0$ to be the case of a CTMC with finite generator matrix $Q$.\footnote{Note that a discrete-time Markov chain $(\pi, P)$ with $P = I + \epsilon Q$, where $\epsilon$ is a small time interval, will tend to the CTMC($\pi, Q$) as $\epsilon \to 0$.}

\begin{table}[hbt]
    \caption{Resulting random walk (RW) or process from the limiting cases of number of vertices $N$ and time interval between transitions $\epsilon$ in the diffusion maps literature \cite{nadler_diffusion_FP_2006}. We highlight the addition of the third entry for CTMCs to complete the set.}

    \label{tab:limits}
    \centering

    \begin{tabular}{p{.15\textwidth}|p{.30\textwidth}|p{.40\textwidth}}
    \hline
    \textbf{Case} & \textbf{Operator} & \textbf{Stochastic Process} \\

    \hline
    $\epsilon > 0$ \newline $N < \infty$ & finite $N\times N$ \newline matrix $P$ & RW\ in discrete space \newline discrete in time (DTMC) \\

    \hline
    $\epsilon > 0$ \newline $N \to \infty$ & operators \newline $T_f, T_b$ & RW\ in continuous space \newline discrete in time \\

    \hline
    $\epsilon \to 0$ \newline $N < \infty$ & infinitesimal generator  \newline matrix $Q \in \mathbb{R}^{N \times N}$ & Markov jump process; discrete in space, continuous in time \\

    \hline
    $\epsilon \to 0$ \newline $N \to \infty$ & infinitesimal \newline generator $\mathcal{H}_f$ & diffusion process \newline continuous in space \& time \\

    \hline
    \end{tabular}
\end{table}

\subsection{Gaussian processes for inferring drift vector field}

Diffusion maps provide a convenient way to embed the CTMC graph into a Euclidean space $E$; however, the push-forward CTMC dynamics is only defined on the image of the embedding, i.e. where the embedded states are. In order to define a fluid approximation, we require a continuous drift vector field to be defined everywhere in $E$. A natural approach is to treat this extension problem as a regression problem, where we use the push-forward dynamics at the isolated state embeddings as observations. We therefore use Gaussian processes (GPs), a non-parametric Bayesian approach to regression, to infer a smooth function $b:E \to \mathbb{R}^d$ that has the appropriate drift vectors where states lie.

A Gaussian process is a collection of random variables $\{f_t\}_{t\in T}$ indexed by a continuous quantity $t$, which follows a distribution over a family of functions $f: T \to \mathbb{R},\ f\in \mathcal{H}$. Over the Hilbert space $\mathcal{H} = L^2(T)$, the distribution can be thought of as an infinite-dimensional Gaussian distribution over function values, where each dimension corresponds to a point on the domain of the function. We write
\[
f(\cdot) \sim \mathcal{GP}(m(\cdot), k(\cdot, \cdot)),
\]
where $m:T \to \mathbb{R}$ is the mean function, and $k: T \times T\to \mathbb{R}$ is the covariance kernel of the distribution over $\mathcal{H}$. The choice of kernel $k(\cdot, \cdot)$ acts as the inner product in the space of functions $\mathcal{H}$, and so determines the kind of functions over which the distribution is defined. Certain kernels define a distribution over a dense subspace of $L^2(T)$, and we therefore say that the GP is a \emph{universal approximator} --- it can approximate any function in $L^2(T)$ arbitrarily well. One such kernel is the squared exponential:
\[
k(t, t') = a^2 \exp(- \frac{\lvert t - t' \rvert^2}{2l^2} ),
\]
where the constants $a$ and $l$ are hyperparameters: the \emph{amplitude} and \emph{lengthscale} respectively.

\paragraph{Gaussian process regression}
Suppose we observe evaluations $\mathbf{f} = (f(t_1), \dots, f(t_n))$ of an (otherwise hidden) function, at points $\mathbf{t} = (t_1,\dots, t_n)$ of the function's domain.
Once an appropriate prior is established, we are able to perform Bayesian inference to obtain a posterior distribution over possible functions consistent with our observations. In Gaussian process regression, the prior is the distribution given by the kernel. The function value $f_\star$ at an unobserved domain point $t_\star$, conditioned on observations $\mathbf{f}$ at points $\mathbf{t}$, follows the predictive distribution:
\[
f_\star \mid \mathbf{f}, \mathbf{t} \sim \int p(f_\star \mid t_\star, \mathbf{t}, \mathbf{f})\ p(\mathbf{f}\mid \mathbf{t})\ d\mathbf{f}.
\]

Since the integral involves only normal distributions, it is tractable and has a closed form solution, which is again a normal distribution. The observations may also be regarded as noisy, which will allow the function to deviate from the observed value in order to avoid extreme fluctuations. Using an appropriate noise model (Gaussian noise), retains the tractability and normality properties. Usually the mean of the predictive distribution is used as a point estimate of the function value. For a comprehensive understanding of Gaussian processes for regression purposes, we refer to \cite{rasmussen_gaussian_2006}.

In our case, the choice of kernel and its hyperparameters is critical, especially when the density of states is low. In the limit of infinite observations of the function, the Gaussian process will converge to the true function over $T$, if the function is in the space defined by the kernel, regardless of the hyperparameters chosen. However, the number of states we embed is finite and so the choice of an appropriate prior can greatly aid the Gaussian process in inferring a good drift vector field. Here, we use the standard squared exponential kernel with a different lengthscale for each dimension, and select hyperparameters which optimise the likelihood of the observations. The optimisation is performed via gradient descent since the gradient for the marginal likelihood is available.

\subsection{The geometric fluid approximation algorithm}
{
Instructions to implementing the geometric fluid approximation are given in Algorithm~\ref{alg:gfa}, detailing the recovery of embedding coordinates using diffusion maps, and inferring the drift vector field using Gaussian process regression. Given initial state coordinates $y(t=0)$ and a duration of time $T$, the inferred drift vector field is used as the gradient in an ODE solver to produce deterministic continuous trajectories in the Euclidean space where states have been embedded. These trajectories are interpreted as approximations to the evolution of the mean of the original process, mapped to the Euclidean embedding space.

\algnewcommand{\LineComment}[1]{\Comment{\parbox[t]{.45\linewidth}{#1}}}
\begin{algorithm}[!htb]

\caption{Geometric fluid approximation (GFA) algorithm. Prototype Python code available at \texttt{https://bitbucket.org/webdrone/gfa/src/master/}.}
\label{alg:gfa}

\begin{algorithmic}[1]

\Procedure{GFA}{$Q \in \mathbb{R}^{N \times N},\ K\in \mathbb{N}_{>0},\ T\in \mathbb{R}_{>0}$,\ $s_0 \in \{1,\dots, N\}$}
  \State Set $\epsilon < \max_i(|Q_{ii}|)$.
  \State $W = D(I + \epsilon Q)$, with $D^{-1} = \text{diag}(I + \epsilon Q)$ a normalising diagonal such that $W_{ii}=1 \forall i$.
  \State $Y = \text{DiffusionMaps}(W, K)$.
  Input $W$ as similarity matrix and $K$ the number of embedding dimensions to \emph{diffusion maps} procedure. Output is $Y \in \mathbb{R}^{N \times K}$, the DM state coordinates.
  \State $R_i = \sum_{j\neq i} (Y_j - Y_i) Q_{ij}\ \forall i \in \{1,\dots, N\}$, where $R \in \mathbb{R}^{N \times K}$ is the vector field values at state embeddings, and single subscripts index matrix rows.
  \State $\tilde{f}(\cdot)| Y, R \sim \mathcal{GP}(\tilde{m}(\cdot), \tilde{k}(\cdot, \cdot))$; GPR for inferring the drift vector field $f: y \mapsto r$ given observations $Y,\ R$.
  \State $(y_t)_0^T = \text{ODE\_solver}(y_0 = Y_{s_0}, dy_t/dt = \langle{\tilde{f}(y_t)}\rangle = \tilde{m}(y_t), t_\text{end} = T)$; produce approximate trajectory (GFA) to the mapped mean trajectory.
\EndProcedure

\Function{DiffusionMaps}{$W,\ K$}
  \State $S = (W + W^\top) / 2$
  \State $\tilde{P} = \text{diag}(S \mathbf{1})$, where $\mathbf{1} = [1,\dots, 1]^\top$.
  \State $V = \tilde{P}^{-1} S \tilde{P}^{-1}$
  \State $\tilde{D} = \text{diag}(V \mathbf{1})$
  \State $H^{(1)}_{ss} = \tilde{D}^{-1 / 2} V \tilde{D}^{-1 / 2}$
  \State $Y = \Phi_{K}$, where $\Phi_K$ is a matrix with columns $(\Phi_1,\dots,\Phi_K)$ the eigenvectors corresponding to the $K$ lowest eigenvalues which solve $H^{(1)}_{ss} \Phi_k = \lambda_k \Phi_k$, excluding the $\lambda=0$ solution.\\
  \Return $Y\in\mathbb{R}^{N\times K}$, the state coordinates.
\EndFunction

\end{algorithmic}
\end{algorithm}
}

{
\subsection{Consistency result}
The \emph{geometric fluid approximation} scheme is applicable in general to all CTMCs; it is therefore natural to ask whether it reduces to the standard fluid approximation on pCTMCs. We have the following result for a related construction, the \emph{unweighted Laplacian fluid approximation}.

\begin{theorem}\label{gfa:thm:consistency}
Let $\mathcal{C}$ be a pCTMC, whose underlying transition graph maps to a multi-dimensional grid graph in Euclidean space using the canonical \emph{hydrodynamic scaling} embedding. The \emph{unweighted Laplacian fluid approximation} of $\mathcal{C}$ coincides with the canonical fluid approximation in the hydrodynamic scaling limit.
\end{theorem}

The proof (see Appendix~\ref{app:grids}) relies on the explicit computation of the spectral decomposition of the Laplacian operator of an unweighted grid graph \cite{klopotek_spectral_2017}, and appeals to the universal approximation property of Gaussian Processes \cite{rasmussen_gaussian_2006}. We conjecture that the conditions for fluid approximation for such a pCTMC will also be satisfied by our \emph{geometric fluid approximation}.

Intuitively, away from the boundaries of the network, the coordinates of the embedded states approach the classical concentration embedding, where each dimension corresponds to a measure of concentration for each species. As the network grows (i.e.\ allowing larger maximum species numbers in the state-space of the chain) states are mapped closer together, reducing jump size, but preserving the ordering. The spacing of states near the centre of the population size is almost regular, approaching the classical density embedding, and the GP smoothing will therefore converge to the classical extended drift field.
}

\section{Empirical observations}

Experimental evidence of our geometric fluid approximation is necessary to give an indication of the method's validity, and a better intuition for its domain of effectiveness. We apply the geometric fluid approximation to a range of CTMCs with differing structure, and present the experimental results in this section. The CTMC models we used are defined in Section~\ref{sec:models}, and the Python code used to produce the results can be found at {\tt https://bitbucket.org/webdrone/gfa/src/master/}.

There is no absolute way to assess whether the method produces a good approximation to the true probability density evolution; we therefore focus on two comparisons: how close the geometric fluid trajectory over time is to the empirical mean of the original CTMC, mapped on the same state-space (Section~\ref{sec:fluid_mean_assess}); and how close the first-passage time (FPT) estimate from the fluid approximation is to the true FPT cumulative density function (estimated by computationally intensive Monte Carlo sampling; Section~\ref{sec:fpt_assess}).

Further, we demonstrate in Section~\ref{sec:subset_mean_assess} how the method is applicable to a subset of the CTMC graph, such that only a neighbourhood of the state-space is embedded. This may result in fluid approximations for graphs whose global structure is not particularly amenable to embedding in a low-dimensional Euclidean space, and so is useful for gauging the behavioural characteristics of the system near a section of the state-space.

In all figures in this section, red lines are solutions of our geometric fluid approximation, obtained via numerical integration of the drift vector field as inferred by GP regression, and blue lines are the mean of CTMC trajectories mapped to the embedding space, which were obtained via Gillespie's exact stochastic simulation algorithm (SSA) \cite{gillespie_exact_1977}. Finally, in figures showing trajectories on the diffusion maps (DM) manifold, grey line intersections are embedded states (the grey lines being the possible transitions, or edges of the network).


\subsection{Models} \label{sec:models}
We examine an array of models to assess the applicability domain of our method. The models are defined below and empirical comparisons for each are presented throughout this section.

\paragraph{Two species birth-death processes}
This model describes two independent birth-death processes for two species, and serves as a basic sanity check. The CTMC graph has a 2D grid structure and in this sense resembles the system in Theorem \ref{gfa:thm:consistency}. In the usual chemical reaction network (CRN) notation, we write:
\begin{equation*}
\emptyset \xrightarrow{10} A,\quad
A \xrightarrow{1/2} \emptyset, \quad
\emptyset \xrightarrow{10} B, \quad
B \xrightarrow{1/2} \emptyset,
\end{equation*}
for the two species $A, B$, and note that, contrary to standard CRN convention for open systems, we introduce a system size variable $N = 30$ such that the count for each species $n_A, n_B$ cannot exceed $N$; this produces a finite-state CTMC that can be spectrally decomposed and embedded. Note further that the birth process involves no particles here, and so transitioning from state $s = (n_A, n_B)$ to state $s' = (n_A + 1, n_B)$ (or from $s = (n_A, n_B)$ to $s' = (n_A, n_B + 1)$) occurs at the same rate of $10 / N$ per second $\forall\ n_A, n_B$. Conversely, death processes are uni-molecular reactions, such that transitioning from $s = (n_A, n_B)$ to $s' = (n_A - 1, n_B)$ occurs at a rate of $(1/2) n_A$ per second $\forall\ n_A, n_B$, as the chemical reaction network interpretation dictates.\footnote{This follows from the definition of the propensity function, see \cite{gardiner_stochastic_2009,schnoerr_approximation_2017} for details.}

\paragraph{Two species Lotka-Volterra model}
This is a Lotka-Volterra model of a predator-prey system. Allowed interactions are prey birth, predators consuming prey and reproducing, and predator death. The interactions with associated reaction rates are defined below in the usual chemical reaction network notation:
\begin{equation*}
R \xrightarrow{b = 1/2} 2R,\quad
R + F \xrightarrow{c = 1/10} 2F,\quad
F \xrightarrow{d = 1/3} \emptyset,
\end{equation*}
where prey is represented by species $R$ (rabbits) and predators by species $F$ (foxes), with maximum predator and prey numbers of $N=30$.

\paragraph{SIRS model}
We describe a widely used stochastic model of disease spread in a fixed population, wherein agents can be in three states: susceptible, infected, and recovered ($S,\ I,\ R$) and a contagious disease spreads from infected individuals to susceptible ones. After some time, infected individuals recover and are immune to the disease, before losing the immunity and re-entering the susceptible state. We define a pCTMC for the process as follows:
\begin{equation*}
S + I \xrightarrow{k_i = 0.1} 2I,\quad
I \xrightarrow{k_r = 0.05} R,\quad
R \xrightarrow{k_s = 0.01} S,
\end{equation*}
where the constants $(k_i,\ k_r,\ k_s)$ have been chosen such that the ODE steady state is reached some time after $t=100$s. The state of the pCTMC at time $t$ is $X(t) = (S(t),\ I(t),\ R(t))$, where $S(t)$ refers to the number of agents in state $S$ at time $t$, and so on for all species.

\paragraph{Genetic switch model}
This is a popular model for the expression of a gene, when the latter switches between two activation modes: \emph{active} and \emph{inactive} \cite{vu_beta-poisson_2016,larsson_genomic_2019}. While active, the gene is transcribed into mRNA at a much faster rate than while inactive (factor of $\sim 10$). The gene switches between the two modes stochastically with a slow rate. We have the following reactions:
\begin{equation*}
P \xleftrightarrow{10^{-4}} \bar{P}\quad
P \xrightarrow{1} A + P,\quad
\bar{P} \xrightarrow{0.1} A + \bar{P},\quad
A \xrightarrow{0.05} \emptyset,
\end{equation*}
where the active and inactive modes are represented by the species $P$ and $\bar{P}$ respectively, with a maximum count of 1. Despite being able to express this model in the usual chemical reaction network language, we emphasize that the binary nature of the switch prohibits usual scaling arguments for reaching the fluid limit.

\subsection{Assessing fluid solution and mean trajectory in embedding space}
\label{sec:fluid_mean_assess}
\begin{figure}[htb]
    \centering
    \includegraphics[width=0.45\textwidth]{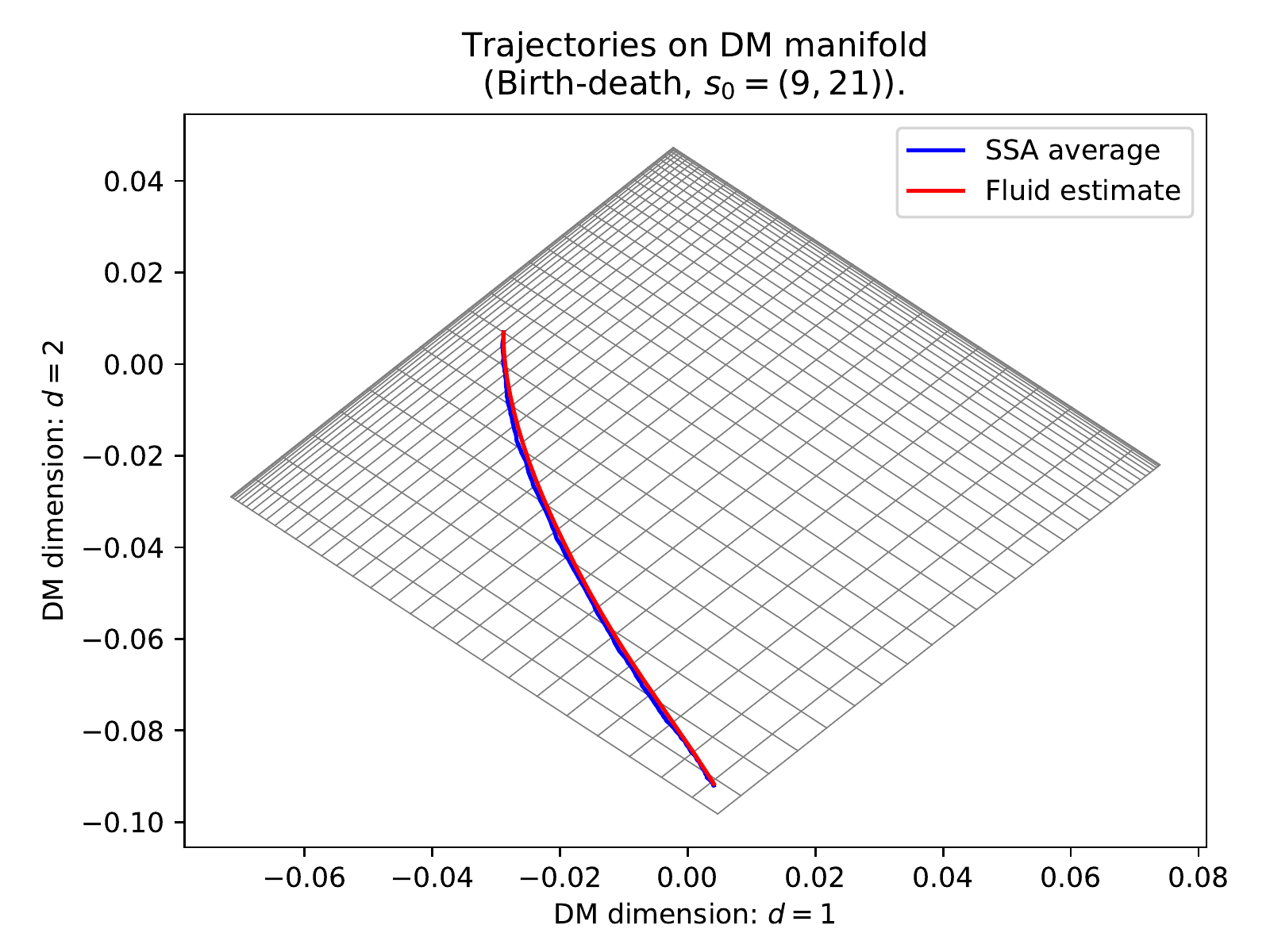}
    ~
    \includegraphics[width=0.45\textwidth]{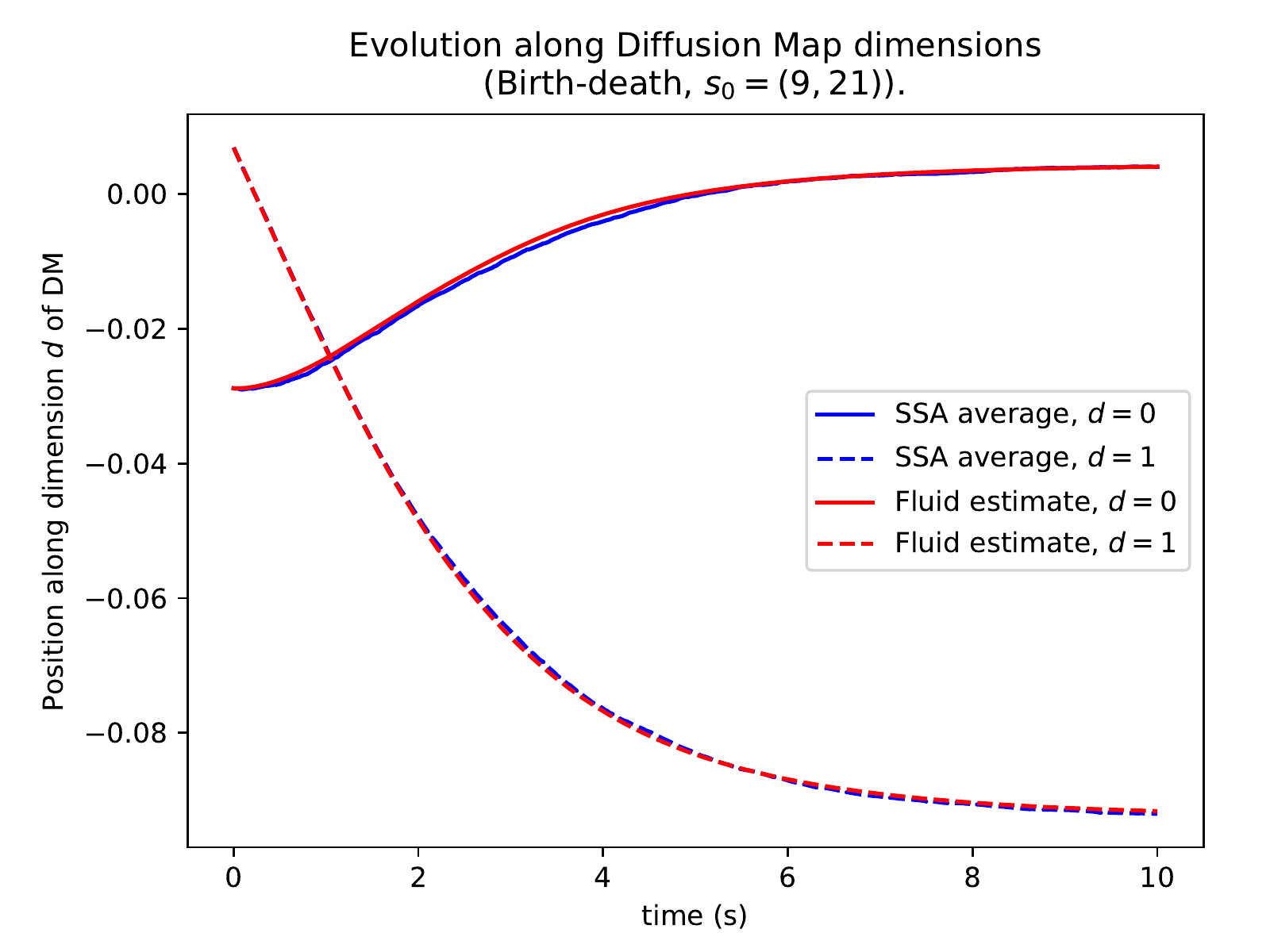}
    \caption{Independent birth-death process for two species, showing the fluid solution (red) and the projected mean evolution (blue). Left: embedded state-space and trajectories in $\mathbb{R}^2$, where grid structure is preserved and species counts are in orthogonal directions. Right: fluid and mean SSA trajectories along embedded dimensions over time.}
    \label{fig:2sp_2d_mid}
\end{figure}

\begin{figure}[htb]
    \centering
    \includegraphics[width=0.45\textwidth]{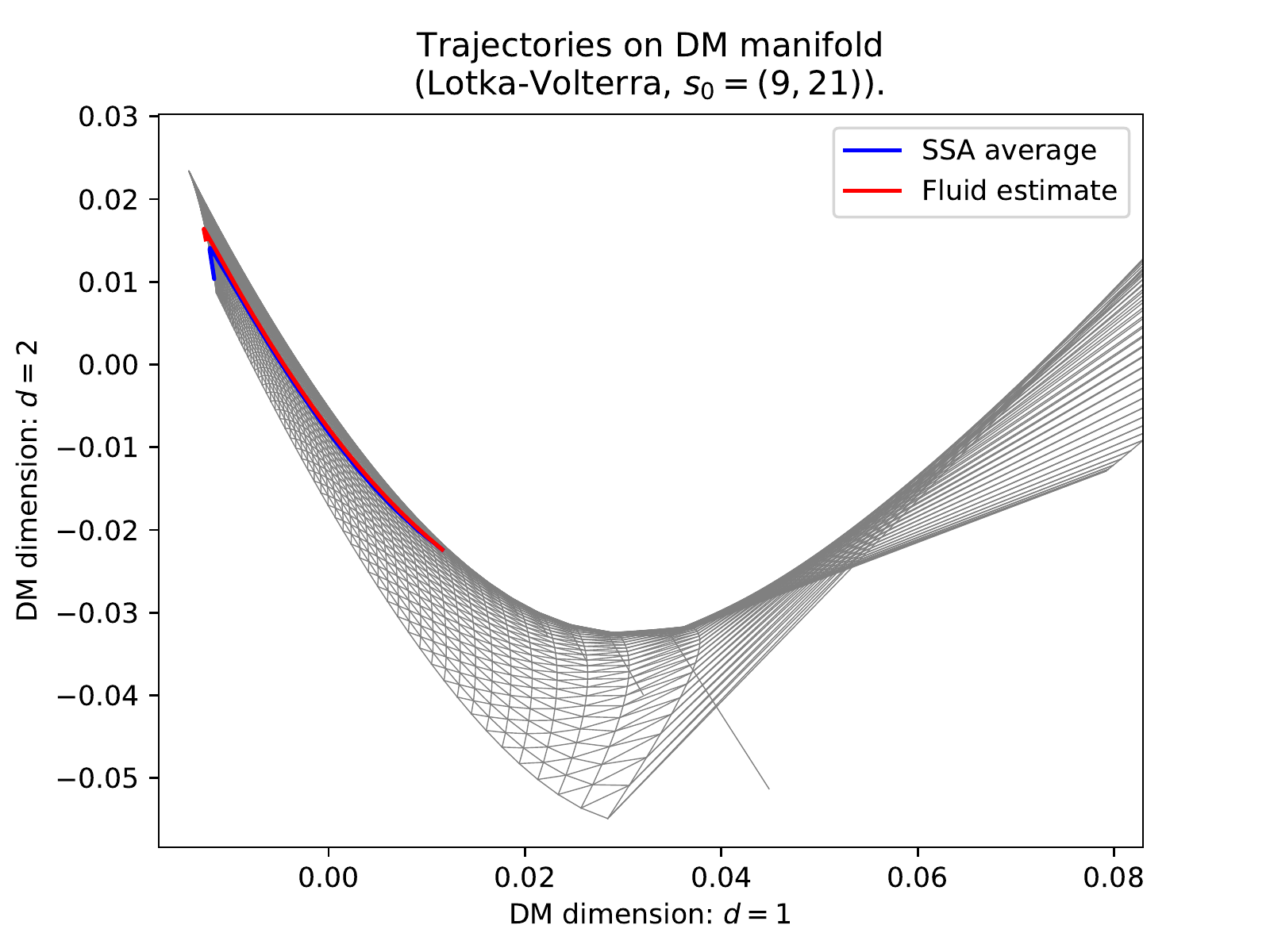}
    ~
    \includegraphics[width=0.45\textwidth]{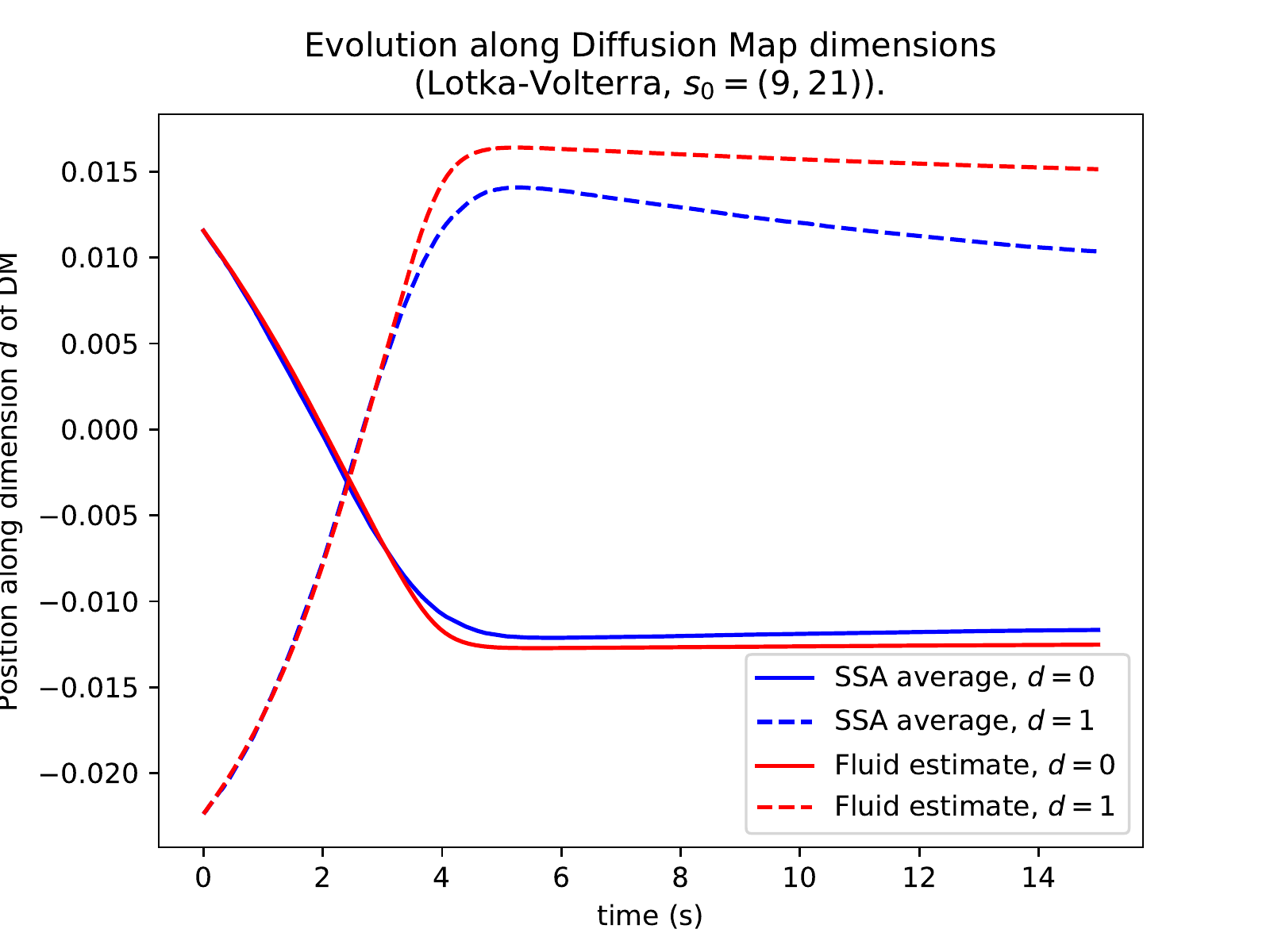}
    \caption{A two species Lotka-Volterra model, showing the fluid solution (red) and the projected mean evolution (blue) slowly diverging from each other. The qualitative behaviour of both is similar as they begin to perform the oscillations typical of this system.}
    \label{fig:2sp_LV}
\end{figure}

\begin{figure}[htb]
\centering
    \begin{subfigure}[t]{0.48\textwidth}
    \includegraphics[width=1.\textwidth]{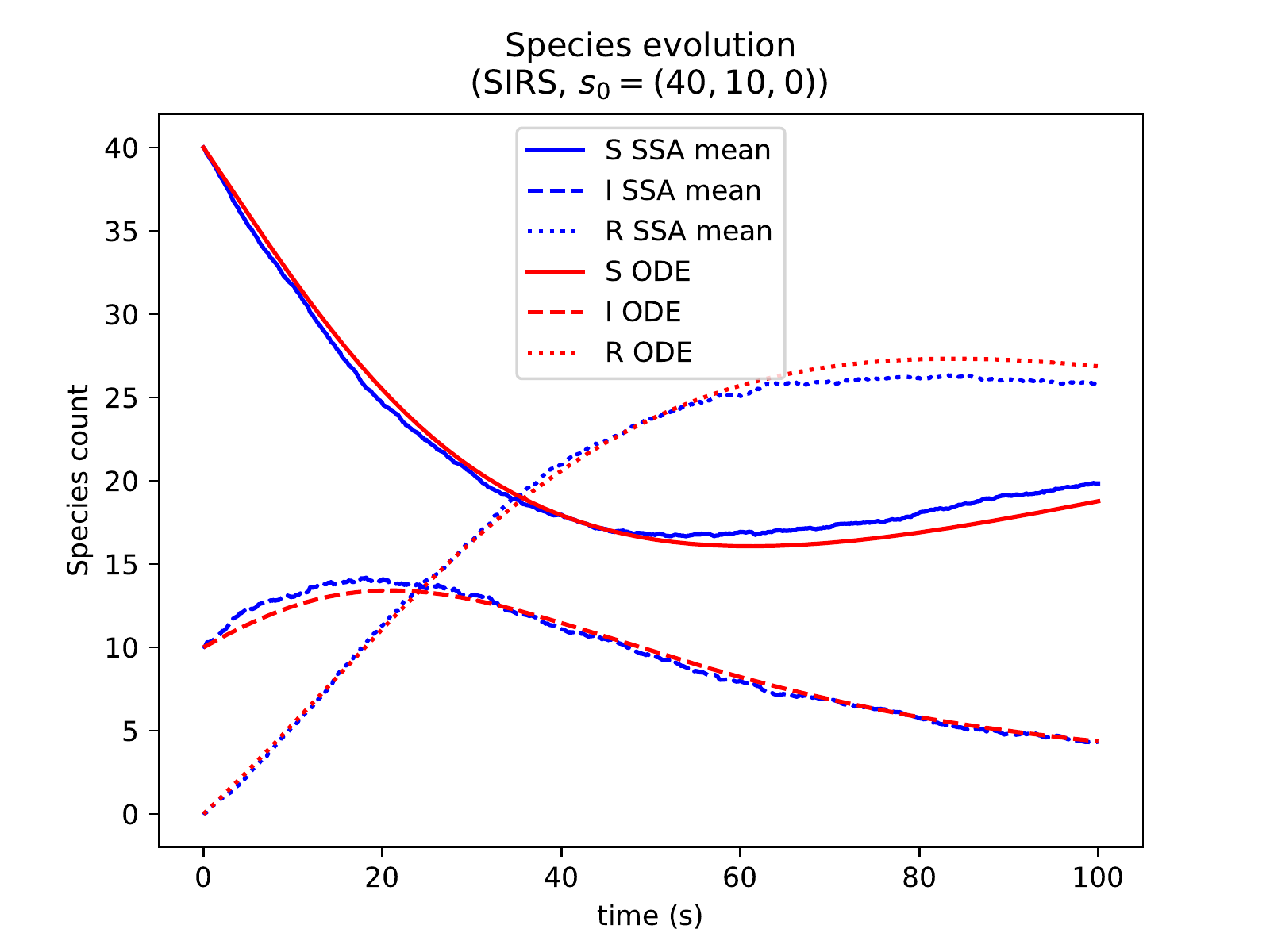}
    \caption{Trajectories of the SIRS model in the space of species counts, each line tracks the count of a species in the system. The classical ODE solution (red) for the three species closely follows the simulation average trajectory (blue).}
    \label{fig:SIR_species}
    \end{subfigure}
\hfill
    \begin{subfigure}[t]{0.48\textwidth}
    \includegraphics[width=1.\textwidth]{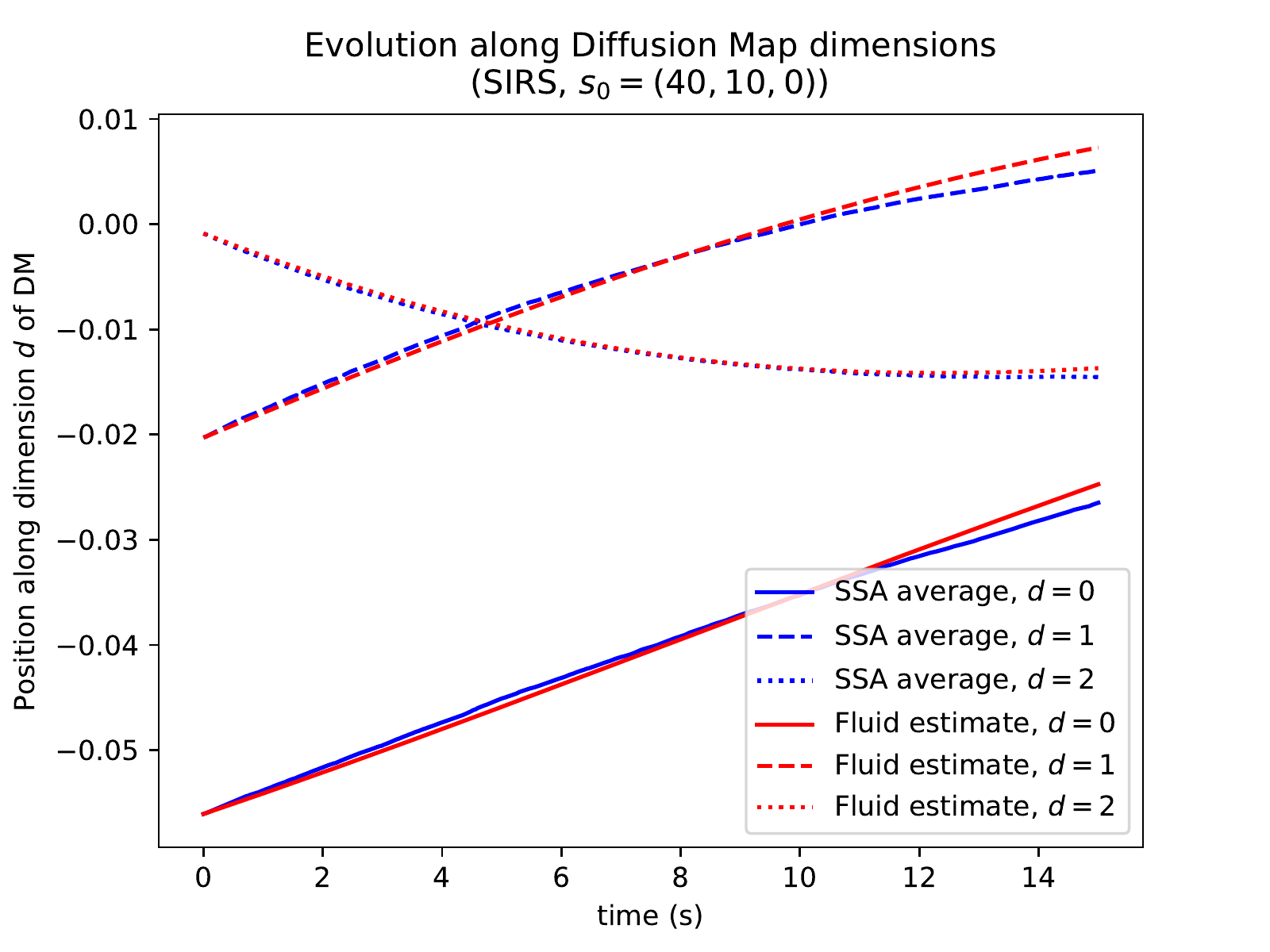}
    \caption{Trajectories of the SIRS model in $\mathbb{R}^3$, where each line tracks the evolution along a dimension $d$ of the DM embedding. The fluid solution (red) closely follows the simulation average trajectory (blue), as in \ref{fig:SIR_species}. Note that these dimensions are no longer interpretable as counts of each species.}
    \label{fig:SIR_3d}
    \end{subfigure}
\caption{Trajectories of the SIRS model with states embedded in continuous space $\subset\mathbb{R}^3$: \ref{fig:SIR_species} the classical embedding to concentration space; \ref{fig:SIR_3d} our embedding with diffusion maps and Gaussian process regression for estimating the drift.}\label{fig:SIR}
\end{figure}

\begin{figure}[htb]
    \centering
    \includegraphics[width=0.45\textwidth]{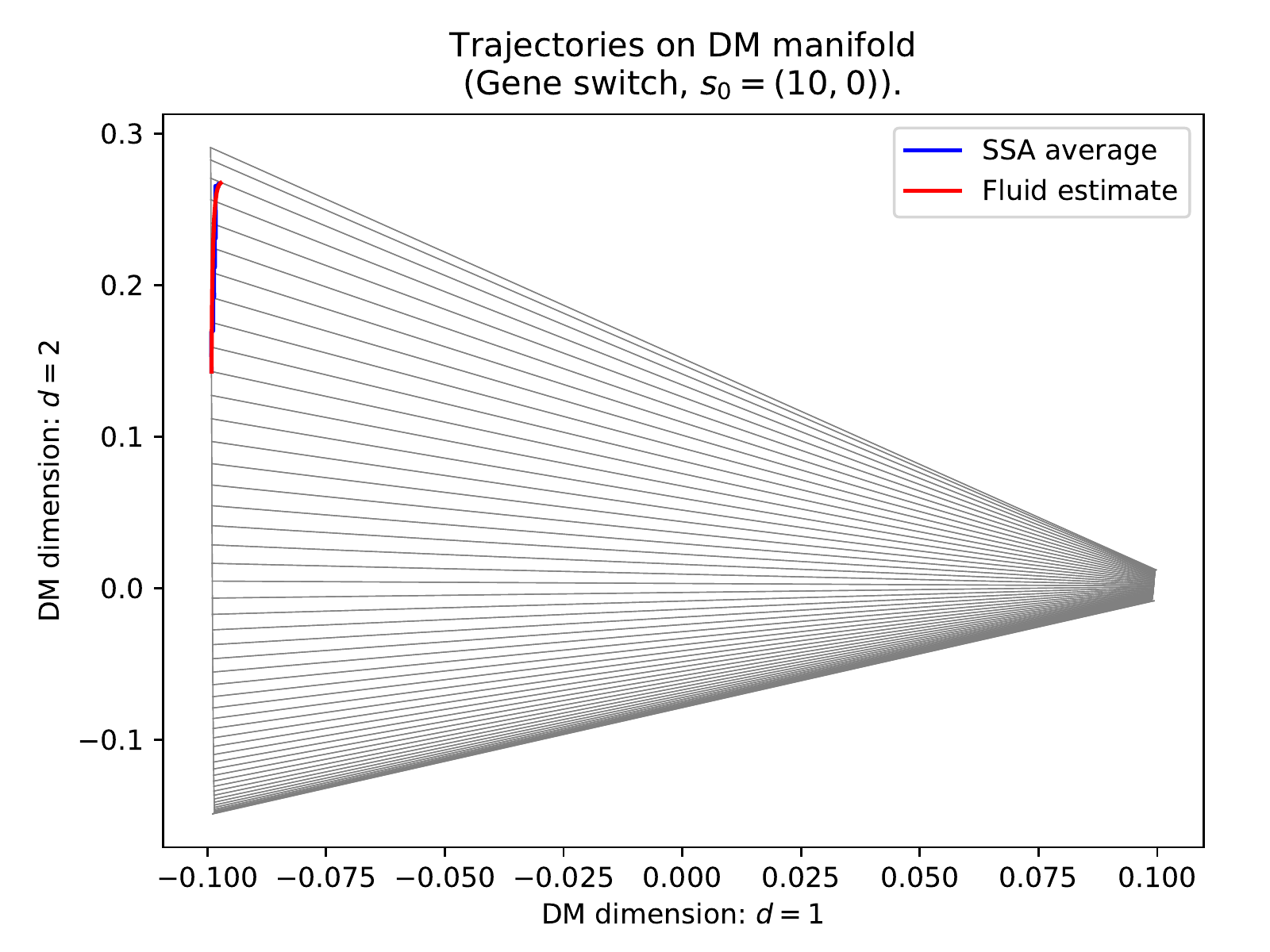}
    ~
    \includegraphics[width=0.45\textwidth]{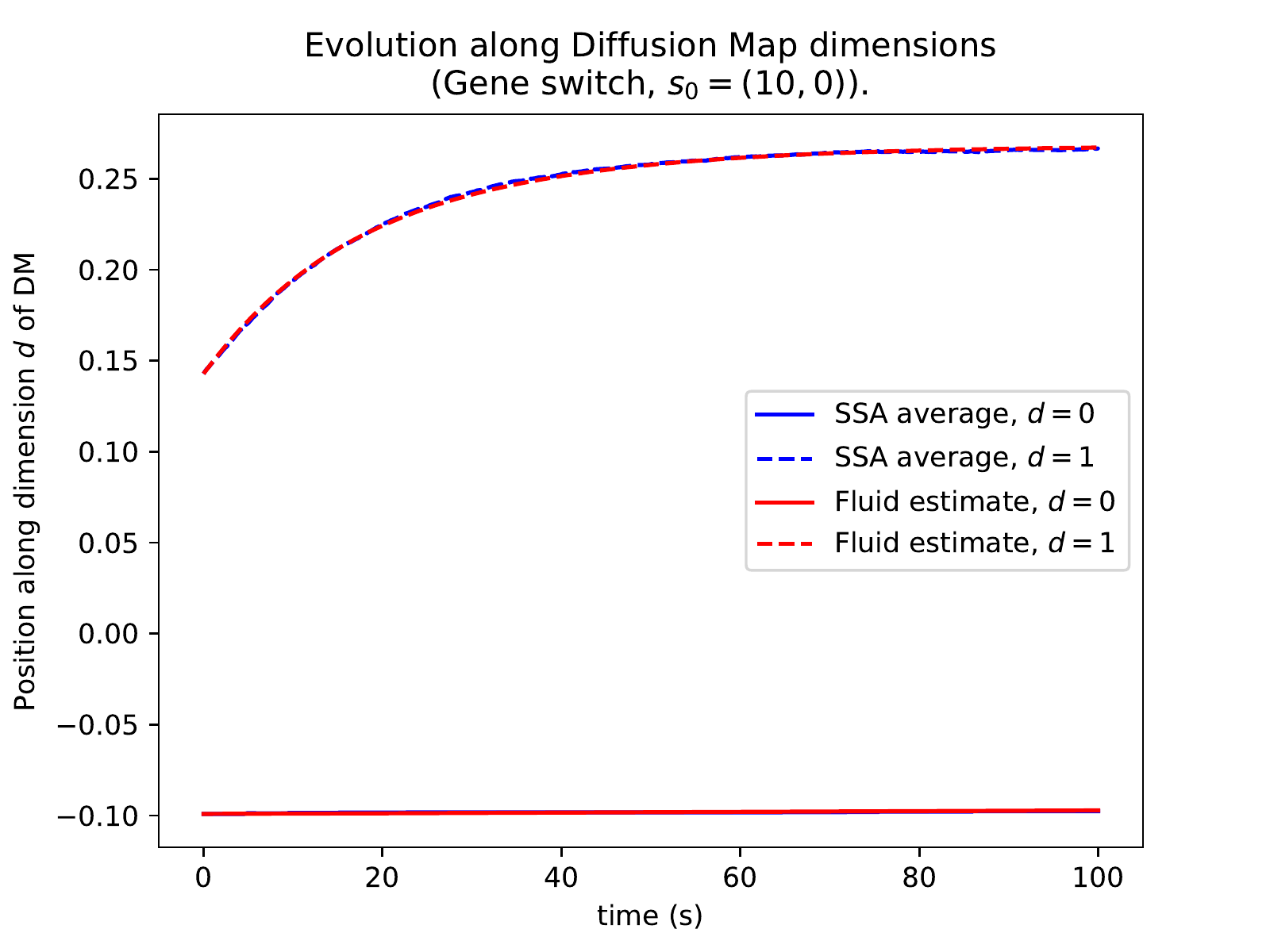}
    \caption{The genetic switch model with switching rate $10^{-4}$s$^{-1}$, showing the fluid solution (red) and the projected mean evolution (blue) keeping close to each other. Transitions from the set of states at $d_1 = -0.1$ (inactive mode) to the set of states at $d_1 = 0.1$ happen very rarely, which is reflected by the mean SSA trajectory.}
    \label{fig:gen_switch_slow}
\end{figure}

\begin{figure}[htb]
    \centering
    \includegraphics[width=0.45\textwidth]{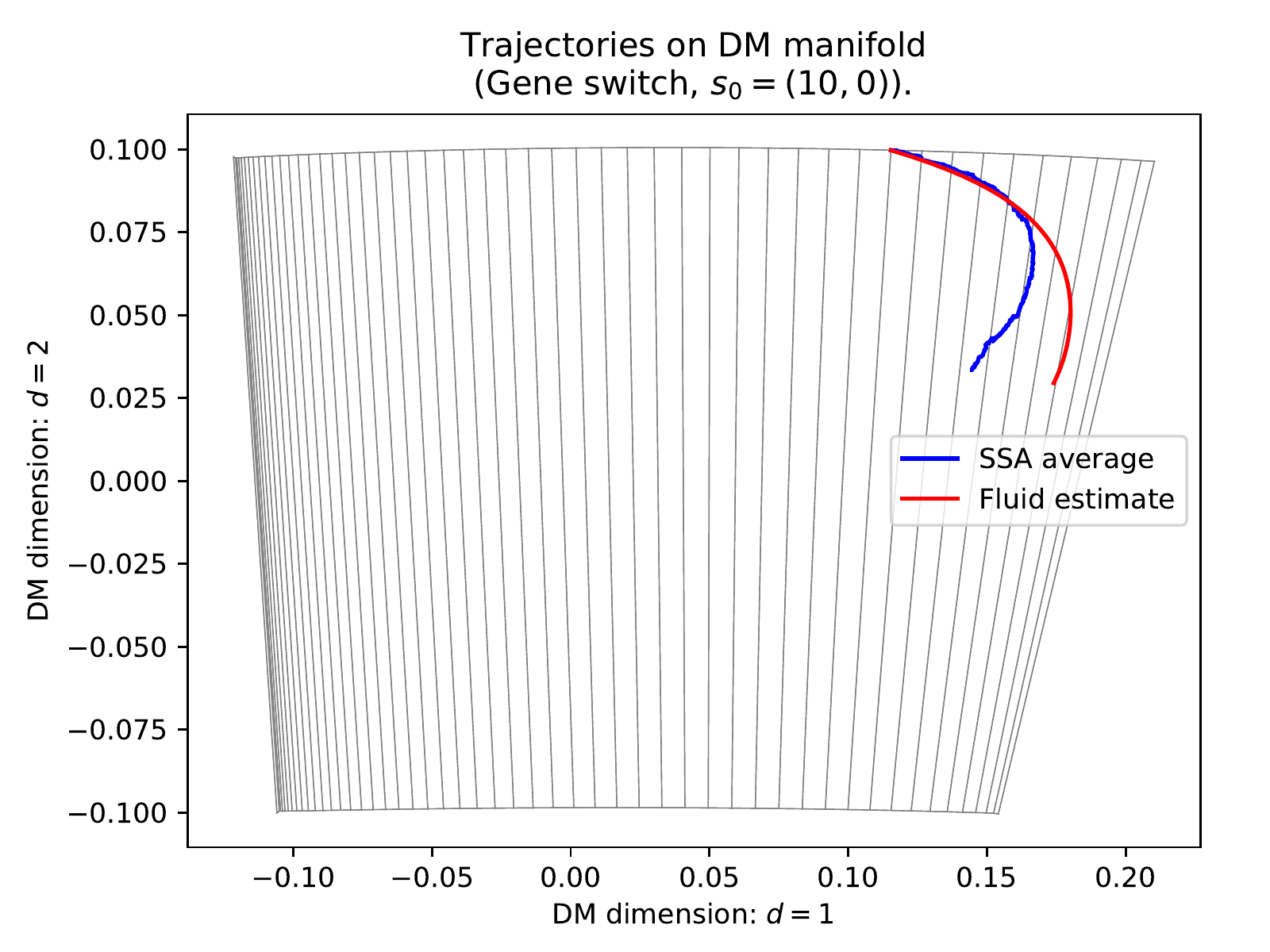}
    ~
    \includegraphics[width=0.45\textwidth]{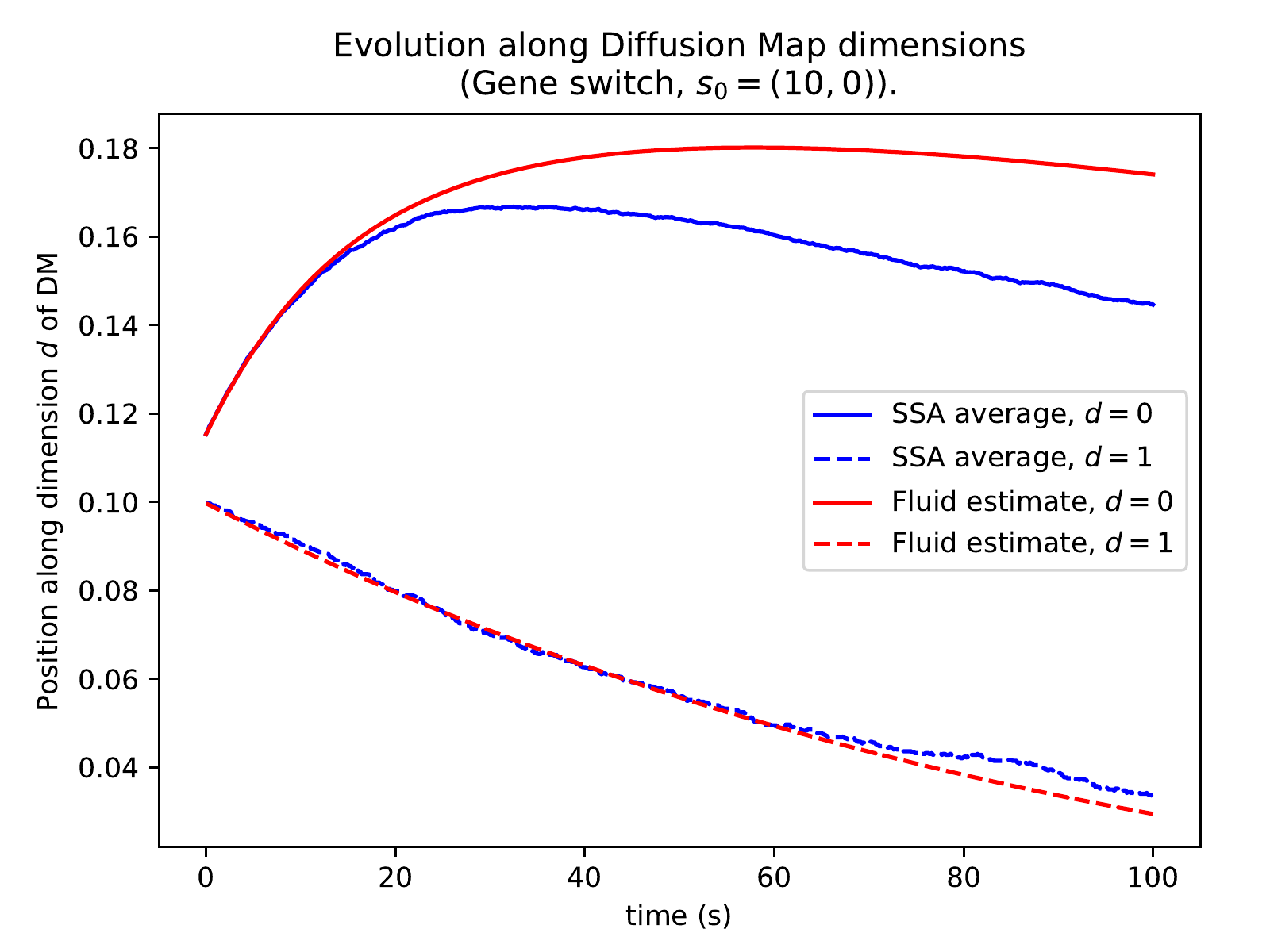}
    \caption{The genetic switch model with a faster switching rate ($5\cdot 10^{-3}$s$^{-1}$), showing how the fluid solution (red) diverges from the projected mean evolution (blue) after $t\approx 20$s; the qualitative aspects of the trajectory remain similar.}
    \label{fig:gen_switch_fast}
\end{figure}

\begin{figure}[htb]
    \centering
    \includegraphics[width=0.45\textwidth]{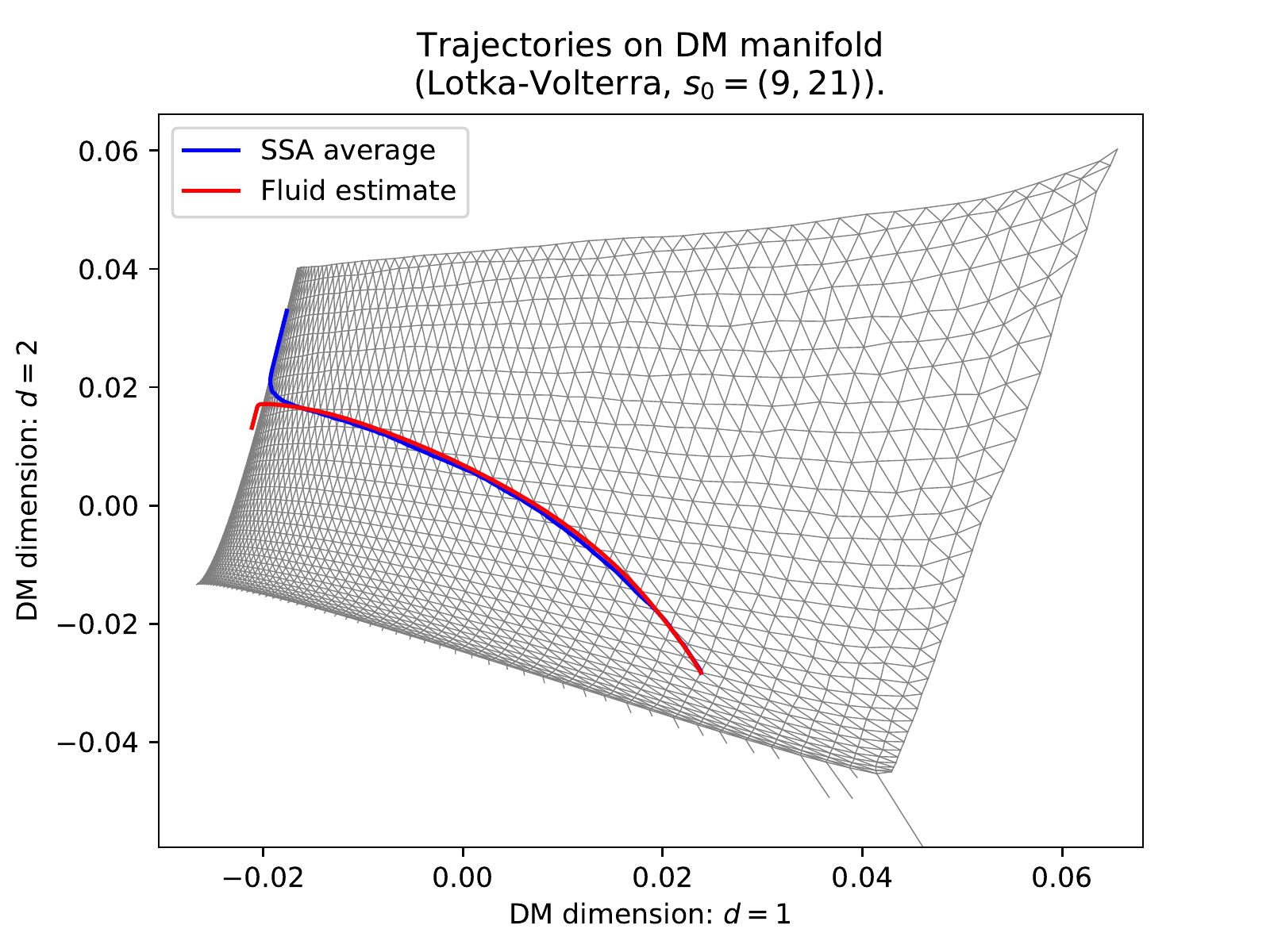}
    ~
    \includegraphics[width=0.45\textwidth]{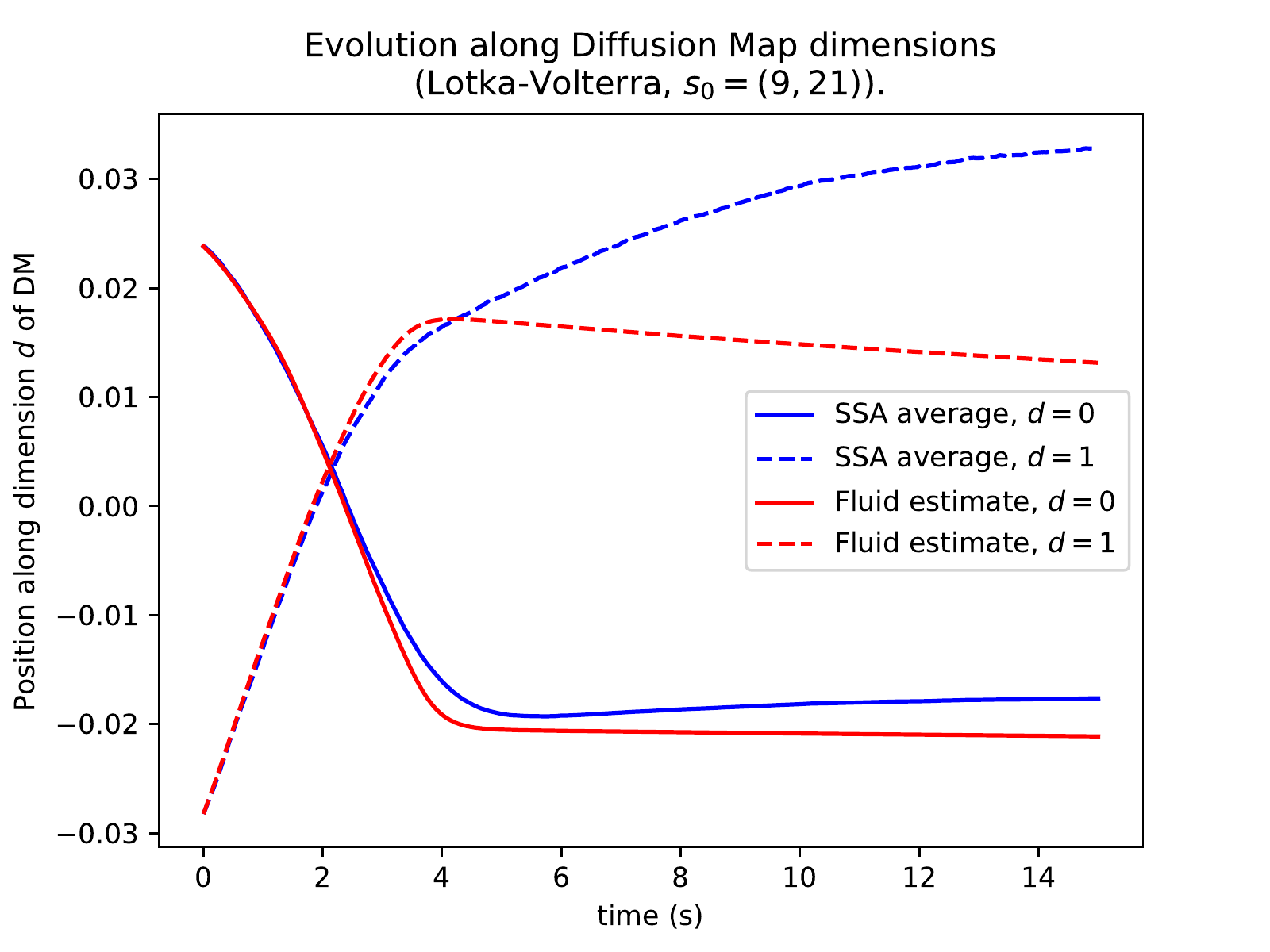}
    \caption{The two species Lotka-Volterra model, with noise added on all transition rates. This is a perturbed pCTMC that is not amenable to classical continuous approximation methods. The fluid solution (red) is close to the projected mean trajectory (blue) away from the boundary.}
    \label{fig:2sp_LV_perturbed}
\end{figure}

\begin{figure}[htb]
    \centering
    \includegraphics[width=0.45\textwidth]{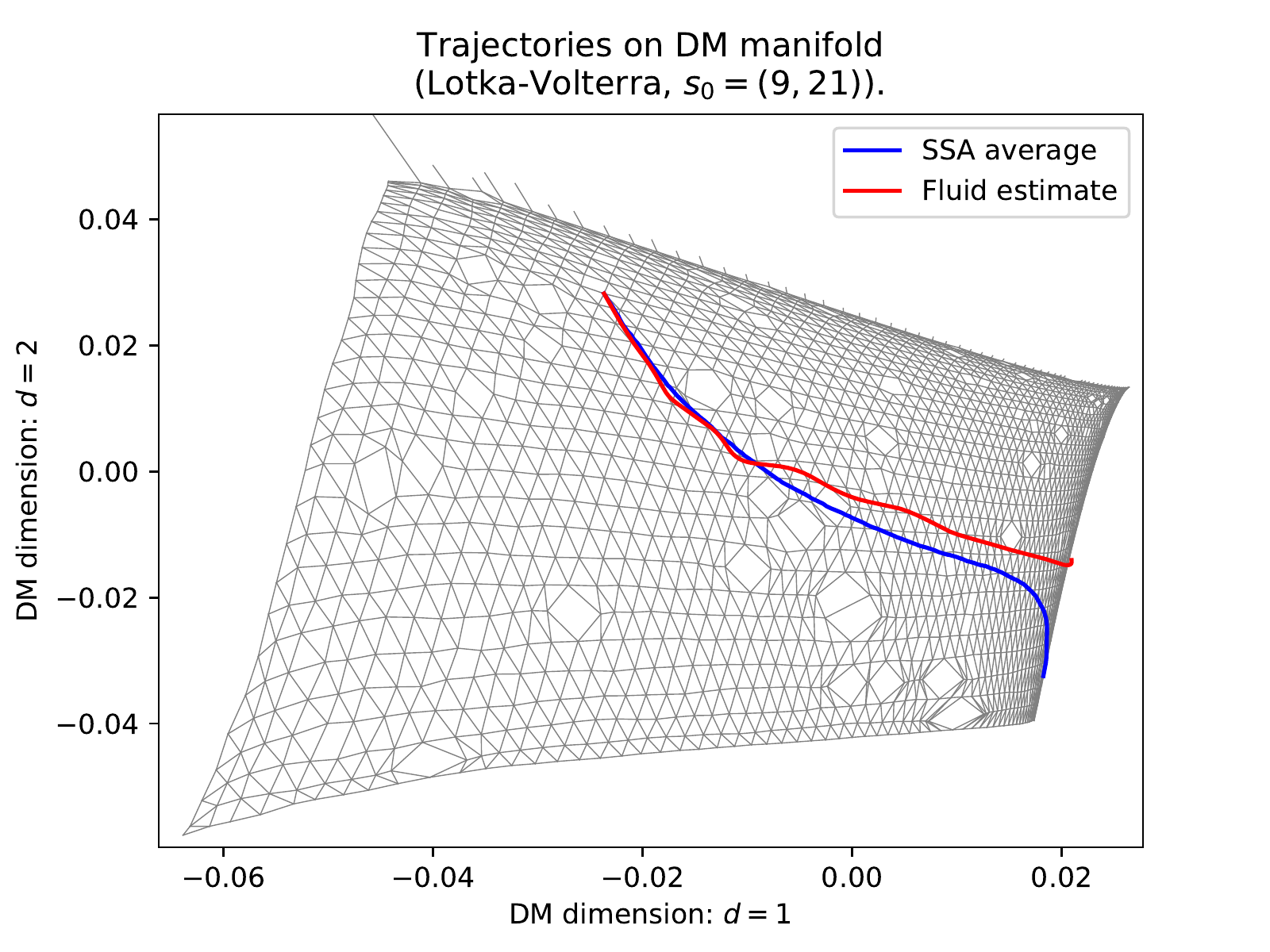}
    ~
    \includegraphics[width=0.45\textwidth]{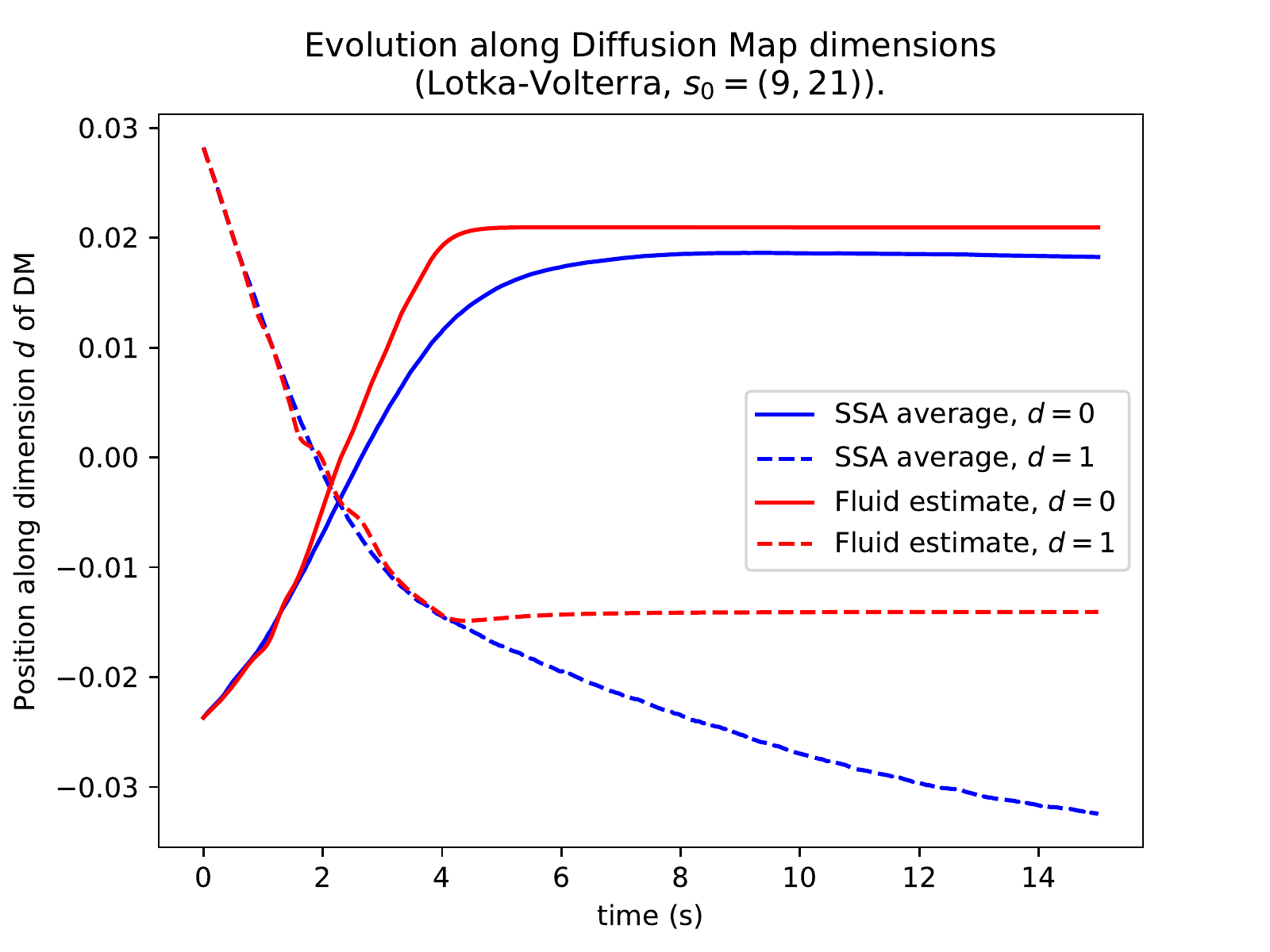}
    \caption{A two species Lotka-Volterra model, perturbed by both noisy transition rates and random removal of transitions. The fluid solution (red) remains similar to the projected mean trajectory (blue) away from the boundary.}
    \label{fig:2sp_LV_perturbed_reduced}
\end{figure}

In our geometric fluid approximation, we create a map using \emph{directed diffusion maps} to embed the CTMC states into a Euclidean space of small dimensionality, and use Gaussian process regression to infer a drift vector field over the space. The resulting continuous trajectories, which we refer to as the \emph{geometric fluid approximation}, are in this section compared to average trajectories of the CTMC systems, projected on the same space.\footnote{Note that in Figures~\ref{fig:2sp_2d_mid}-\ref{fig:2sp_LV_subset} when referring to `dimension $d=j$' of the diffusion map projection, we refer to the $j$th coordinate of the embedding of the manifold in Euclidean space, as recovered by diffusion maps. } The latter are obtained by drawing 1000 trajectories of the CTMC using the SSA algorithm, and taking a weighted average of the state positions in the embedding space.

Our geometric fluid approximation does well for pCTMC models, where we know that the state-space can be naturally embedded in a Euclidean manifold. This is especially true for systems like the independent birth-death processes of two species, which do not involve heavy asymmetries in the graph structure. The more the structure deviates from a pCTMC and the more asymmetries in the structure, the larger the deviations we expect from the mean SSA trajectory. Additionally, we expect large deviations in the case of bi-modal distributions over the state-space, as is the general case for fluid approximations. This is because the latter are point-mass approximations of a distribution, and so are naturally more suited to approximate uni-modal, concentrated densities.

\paragraph{Two species birth-death processes}
As a sanity check, we examine how our method approximates the mean trajectory of the trivial system of two independent birth-death processes described above. The true distribution for such a system is uni-modal in the usual concentration space, and the graph has the structure of a 2D grid lattice with no asymmetries. As shown in Figure~\ref{fig:2sp_2d_mid}, the geometric fluid approximation is very close to the empirical mean trajectory, which supports our consistency theorem and expectations for agreement in the case of symmetric graphs.

\paragraph{Lotka-Volterra model}
We perform our geometric fluid approximation for the non-trivial case of a Lotka-Volterra system, which models a closed predator-prey system as described above. The asymmetric consumption reaction distorts the grid structure representative of the Euclidean square two species space. Therefore, the manifold recovered is the Euclidean square with shrinkage along the consumption dimension --- more shrinkage is observed where predators and prey numbers are higher, since this implies faster consumption reactions. We observe in Figure~\ref{fig:2sp_LV} that the fluid estimate keeps close to the mean initially and slowly diverges; however, the qualitative characteristics of the trajectory remain similar.

\paragraph{SIRS model}
The SIRS model gives us the opportunity to compare trajectories in the embedding space of the geometric fluid, with trajectories in the concentration space used by the standard fluid approximation. We observe in Figure~\ref{fig:SIR} good agreement with the empirical mean trajectory for both fluid methods.

The classical fluid trajectory (Figure~\ref{fig:SIR_species}) is attainable in terms of the concentration of each species; it evolves according to coupled ODEs:
\begin{equation}
\frac{ds}{dt} = k_s r(t) -\frac{k_i}{N} i(t) s(t), \quad
\frac{di}{dt} = \frac{k_i}{N} i(t) s(t) - k_r i(t), \quad
\frac{dr}{dt} = k_r i(t) - k_s r(t),
\end{equation}
where $x(t) = (s(t),\ i(t),\ r(t)) = (S(t),\ I(t),\ R(t)) / N$, and $N \in \mathbb{N}_{>0}$ is the total population. Increasing $N$ linearly scales the ODE solution without affecting the dynamics; the SSA average converges to the ODE solution as $N\to\infty$. Similarly, Figure~\ref{fig:SIR_3d} shows the fluid solution in $\mathbb{R}^3$ obtained by our geometric fluid approximation.

\paragraph{Genetic switch model}
The model of a genetic switch is a departure from the usual pCTMC structure, since the binary switch introduces very slow mixing between two birth-death processes each with a different fixed point. The bi-modality of the resulting steady-state distribution is problematic to capture for any point-mass trajectory, and quickly leads to divergence of the fluid trajectory from the mean. With the particularly slow switching rate of $10^{-4}$s$^{-1}$, our method produces fluid trajectories close to the mean trajectory for up to $100$s, mostly because the mixing is very slow and the distribution remains relatively concentrated for a long time (Figure~\ref{fig:gen_switch_slow}). However, with the faster rate of $5\cdot 10^{-3}$s$^{-1}$, our fluid approximation quickly diverges from the mean trajectory (Figure~\ref{fig:gen_switch_fast}), as the expected result of faster mixing.

\paragraph{pCTMC perturbations} 
It is expected that the method will perform well for CTMCs that are in some sense similar to a pCTMC, but cannot be exactly described by a chemical reaction network. We therefore demonstrate how the method performs for perturbations of a Lotka-Volterra system. To achieve the perturbation, we add noise to every existing transition rate (non-zero element of $Q$) of the Lotka-Volterra system we had above. The perturbed transition matrix $Q_{\text{per}}$ is described in terms of the Lotka-Volterra matrix $Q_{LV}$ by
\begin{align}
[Q_\text{per}]_{ij} = \begin{cases}
                        [Q_{LV}]_{ij} + |\eta_{ij}|, &\text{if}\ [Q_{LV}]_{ij} > 0,\\
                        0 & \text{otherwise},
                    \end{cases}
\end{align}
for all $i\neq j$, where $\eta_{ij} \sim \mathcal{N}(0, 0.5^2)$, and $[Q_\text{per}]_{ii} = \sum_j [Q_\text{per}]_{ij}$ as usual. The projection in Figure~\ref{fig:2sp_LV_perturbed} shows that our method performs reasonably well near the pCTMC regime, where no classical continuous state-space approximation method exists.

A different kind of perturbation is achieved by randomly removing possible transitions of the original pCTMC. This amounts to setting some off-diagonal elements of the $Q$ matrix to 0, and re-adjusting the diagonal so that all rows sum to 0. In order to avoid creating absorbing states or isolated states, we remove transitions randomly with a probability of 0.1. Our method performs reasonably under both kinds of perturbations, as seen in Figure~\ref{fig:2sp_LV_perturbed_reduced}.

\subsection{Embedding a subset of the system}
\label{sec:subset_mean_assess}
The empirical success of the method on perturbed pCTMC systems encouraged further exploration in cases where there is no global continuous approximation method, but the CTMC graph has regions which resemble a pCTMC structure, or are otherwise suitable for embedding in a continuous space. Consequently, we sought to embed only a subset of the state-space of a CTMC. Embedding state-space subsets can be useful for CTMCs that have a particularly disordered global structure (e.g.\ require many dimensions, or have areas on the manifold with low density), but which may contain a neighbourhood of the state-space that better admits a natural embedding. Additionally, one could introduce coffin states near the boundary of a pCTMC to apply the method on reachability problems.

{
A subset includes every reachable state within $r$ transitions from a selected \emph{root state} $s_r$, denoted as $\Delta(s_r, r)$. Transitions from or to states outside the selected subset are ignored, and the remaining $Q$ matrix is embedded in $\mathbb{R}^2$. The drift vectors on boundary states lack all components of transitions outside the subset, and so the probability flux is inaccurate on the boundary. Figure~\ref{fig:2sp_LV_subset} shows the Lotka-Volterra model subset $\Delta(s_r=(R=5, F=9), r=8)$, embedded in $\mathbb{R}^2$. We can see that the behaviour near the root state is close to the projected sample mean evolution, despite the boundary issues.

\begin{figure}[htb]
    \centering
    \includegraphics[width=0.45\textwidth]{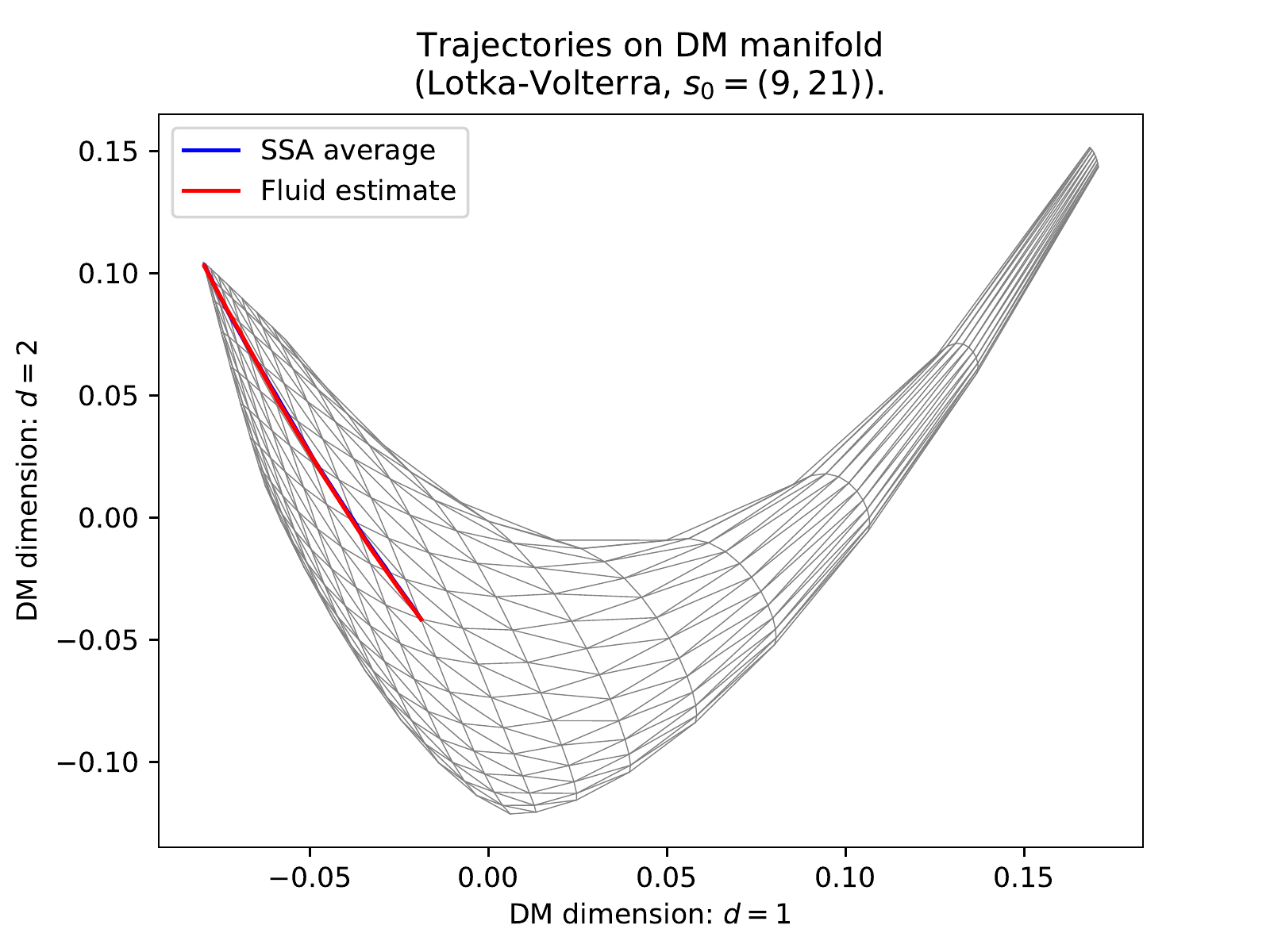}
    ~
    \includegraphics[width=0.45\textwidth]{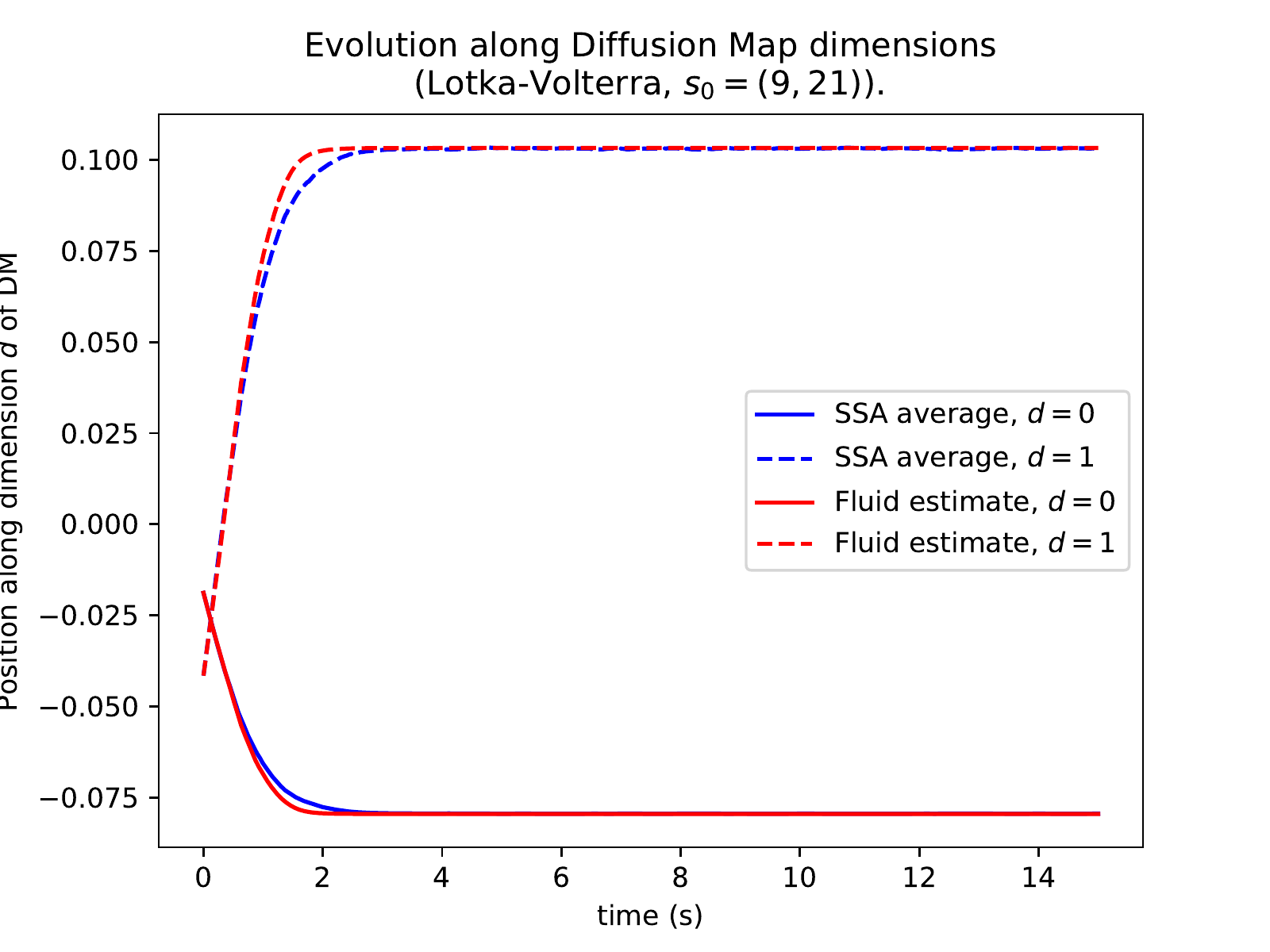}
    \caption{Embedding the subset $\Delta(s_r=(R=5, F=9), r=8)$ of the two species Lotka-Volterra model. The fluid solution (red) remains similar to the projected mean trajectory (blue) away from the boundary, despite the boundary inaccuracies of the probability flux.}
    \label{fig:2sp_LV_subset}
\end{figure}
}


\subsection{First passage times}
\label{sec:fpt_assess}
Another common quest of such approximation techniques is estimating the \emph{first passage time} (FPT) distribution for a target subset of states of the Markov chain. Literature on this is rich --- there has been significant effort in this direction, utilising both established probability evolution methods and constructing new theoretical methods tailored to this problem \cite{darling_first_1953,hayden_fluid_2012,schnoerr_efficient_2017}. The former is possible since FPT estimation can be formulated as the classical problem of estimating how the probability distribution over the state-space evolves for a modified version of the Markov chain in question.

Specifically, consider a Markov chain with rate matrix $Q$ for the state-space $I$. Let $B \subseteq I$ be a set of target states for which we want to estimate the distribution for the FPT $\tau$, given some initial state $\xi_0 \in I \setminus B$. The FPT cumulative density function (CDF) is equivalent to the probability mass on the set $B$ at time $\tau$, if every state in $B$ is made absorbing. In this manner, many methods for approximating probability density evolution over the state-space of a CTMC can also be used to approximate FPT distributions.

\paragraph{The fluid proximity approach}
A natural avenue to estimate the FPT when a fluid approximation to the CTMC exists, is to consider how close the fluid solution is to the target set $B$. The classical fluid approximation usually relies on population structured CTMCs, where the target set is often a result of some population ratio threshold (e.g.\ all states where more than 30\% of the total population is of species $A$: $N_A / N > 0.3$). Since the set is defined in terms of population ratios, it is trivial to map threshold ratios to the continuous concentration space where the pCTMC is embedded, and hence define corresponding concentration regions. The time at which the fluid ODE solution enters that region of concentration space is then an approximation for the FPT CDF. The latter will of course be a step function (from 0 to 1) since the solution is the trajectory of a point mass. Keeping the same threshold ratios for the target set, and scaling the population size $N$ should drive the true FPT CDF towards the fluid approximation. If more moments of the probability distribution are approximated (for instance in moment closure methods) one can derive bounds for the FPT CDF; these can be made tighter as higher order moments are considered, as shown in \cite{hayden_fluid_2012}.

In our case, the fluid ODE solution only tracks the first moment of the distribution which implies a point mass approximation. Additionally, we have done away with the population structure requirement, such that thresholds for defining target sets are no longer trivially projected to the continuous space where we embed the chain. The latter challenge is overcome by considering the Voronoi tessellation of the continuous space, where each embedded state serves as the \emph{seed} for a Voronoi cell. We then say that the fluid solution has entered the target region if it has entered a cell whose seed state belongs in the target set $B$. Equivalently, the solution is in the target region when it is closer (with Euclidean distance as the metric) to any target state than to any non-target state.

Checking which is the closest state is computationally cheap, and so we can produce FPT estimates at little further cost from the fluid construction.
{
Results for the SIRS model, the Lotka-Volterra and perturbed Lotka-Volterra models follow.
}

\begin{figure}[htb]
\centering
\includegraphics[width=0.7\textwidth]{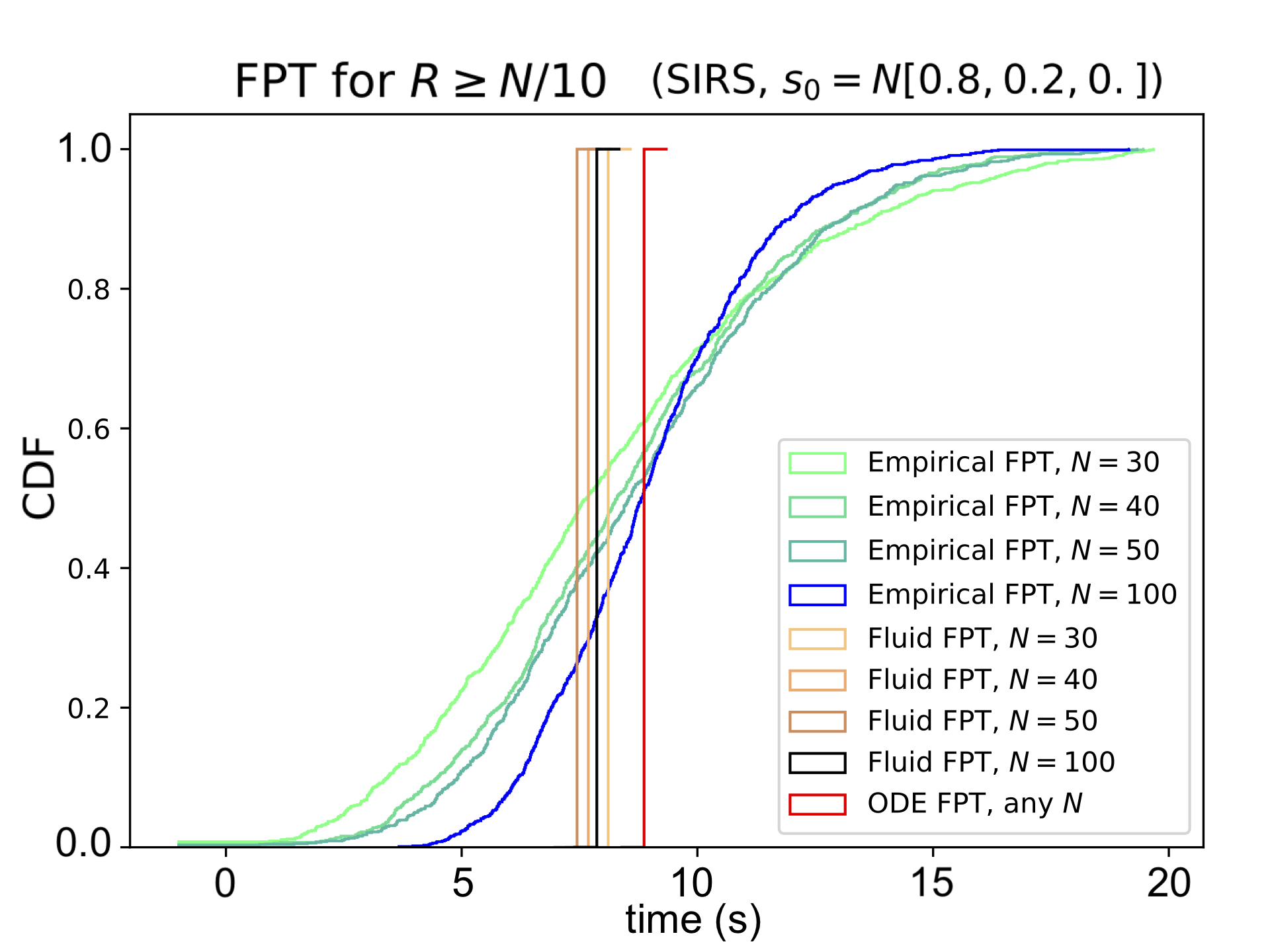}
\caption{First passage time CDFs for the SIRS model with different populations. The classical solution gives the same estimate for all $N$, to which the SSA estimates converge as $N\to\infty$. Naturally, both the classical and our estimates are single step functions, since we approximate the probability distribution evolution by a point mass. We are consistently close to both the SSA and classical fluid CDFs.}
\label{fig:SIR_3d_fpt}
\end{figure}

\paragraph{FPT in the SIRS model}
We define a set of \emph{barrier states} in the SIRS model, $B = \{(S,\ I,\ R) \mid R/N\geq 1 / 10\}$, and examine the FPT distribution of the system into the set $B$, with initial state $X(0) \notin B$. Note that the trivial scaling laws for this model, owing to the fixed population size, makes it simple to identify corresponding barrier regions in concentration space: $b = \{(s,\ i,\ r) \mid r\geq 1 / 10\}$. We can therefore compare the fluid solution FPT estimate to the empirical CDF (trajectories drawn by the SSA), as well as to our own fluid construction with an embedding given by diffusion maps and a drift vector field estimated via a Gaussian process.  Figure~\ref{fig:SIR_3d_fpt} shows that our approach is in good agreement with both the empirical mean FPT and the classical fluid result.

\paragraph{FPT in the Lokta-Volterra model}
Here we embed the Lotka-Volterra model, and define the barrier set of states $B = \{(R, F) \mid 0.6 N > F \geq 0.2 N\}$ for which we estimate FPT CDFs, with initial state $X(0) = (0.3, 0.7) N$, for various system sizes $N=\{30,40,50\}$.

\begin{figure}[htb]
\centering
\includegraphics[width=0.45\textwidth]{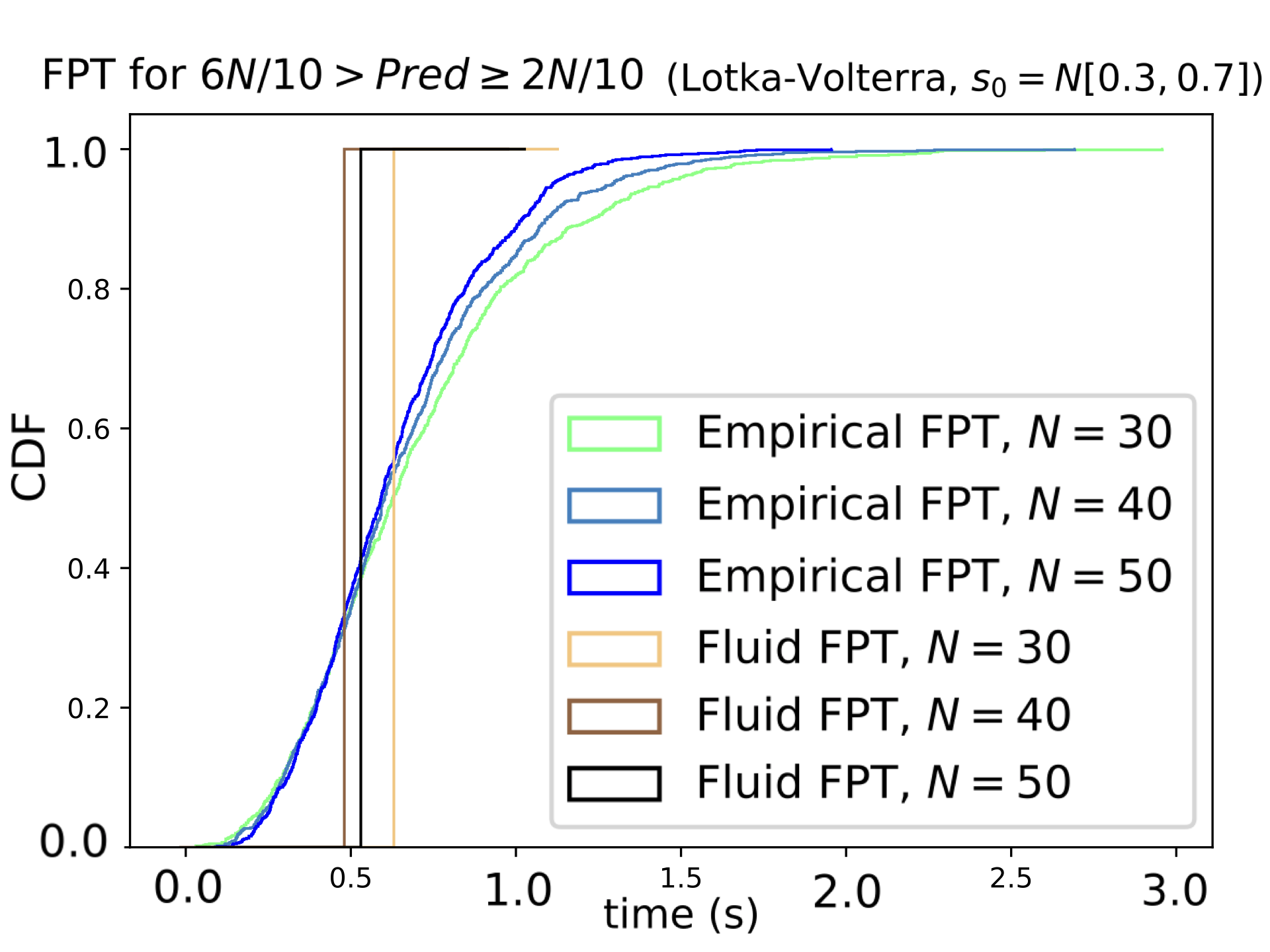}
~
\includegraphics[width=0.45\textwidth]{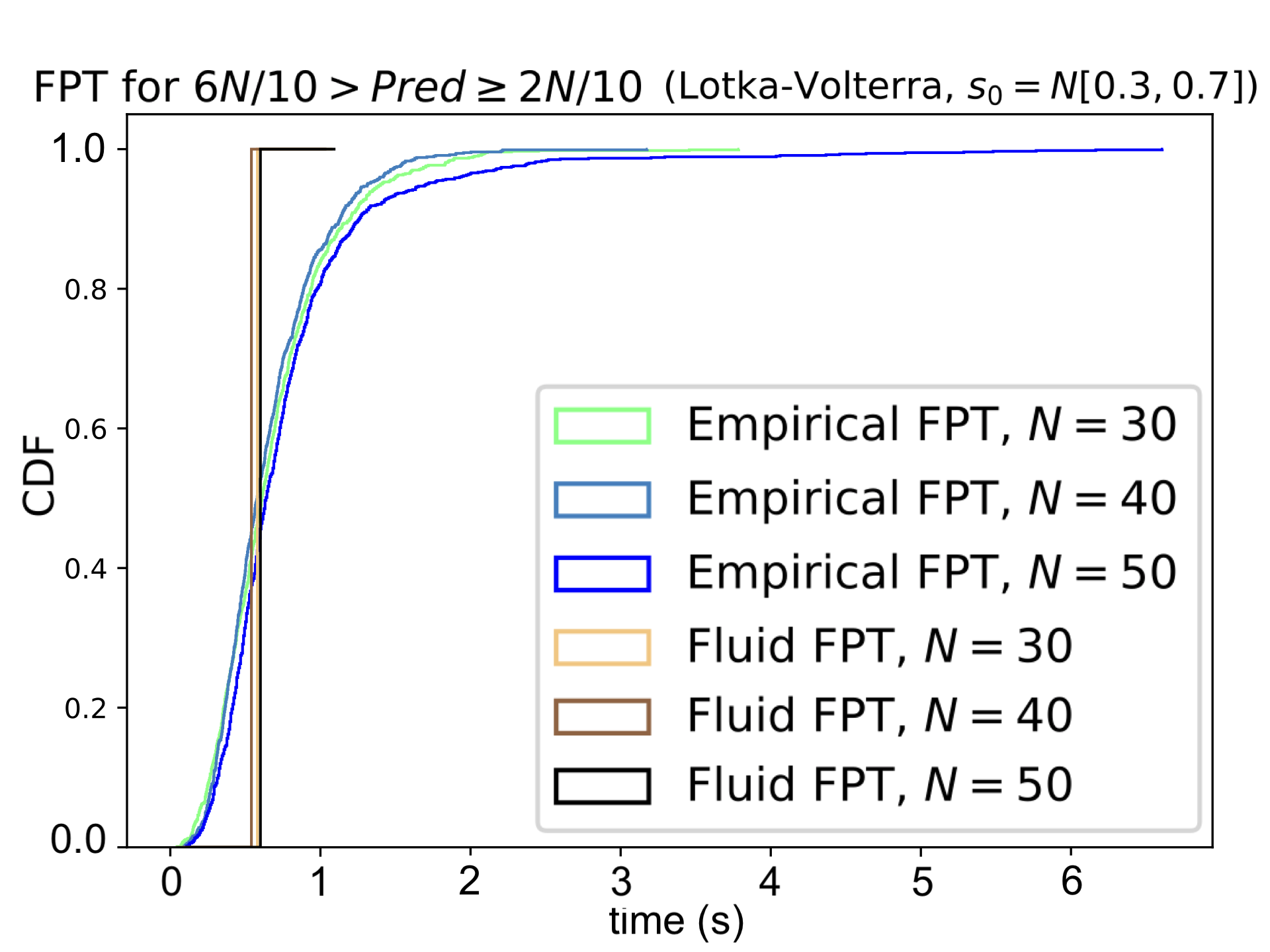}
\caption{
First passage time CDFs for the Lotka-Volterra model, with $B=\{(R, F) \mid 0.6 N > F \geq 0.2 N\}$. Left: unperturbed LV model; the fluid CDF step function crosses the SSA CDF at $\sim0.4-0.5$, which is a reasonable estimate for a point mass approximation. Right: LV model perturbed by both noisy transition rates and random removal of transitions; the fluid CDF estimate is consistently close to the SSA CDF as the system size $N$ increases.}
\label{fig:LV_fpt}
\end{figure}

We show in Figure~\ref{fig:LV_fpt} (left) how our fluid construction estimates an FPT close to the SSA CDF. This is expected when embedding a structured model such as the Lotka-Volterra, where two dimensions are adequate to preserve the network topology and the Gaussian process can well approximate the continuous drift vector field. Finally, we show in Figure~\ref{fig:LV_fpt} (right) that a good estimate of the FPT is recovered for the perturbed Lotka-Volterra, which is no longer a chemical reaction network.

\section{Conclusions}
CTMCs retain a central role as models of stochastic behaviour across a number of scientific and engineering disciplines. For pCTMCs, model approximation techniques such as fluid approximations have played a central role in enabling scalable analysis of such models. These approximations, however, critically rely on structural features of pCTMCs which are not shared by general CTMCs. In this paper, we presented a novel construction based on machine learning which extends fluid approximation techniques to general CTMCs. Our new construction, the \emph{geometric fluid approximation}, is {
(with certain hyperparameters)} equivalent to classical fluid approximations for a class of pCTMCs; empirically, the geometric fluid approximation provides good quality approximations in a number of non-trivial case studies from epidemiology, ecology and systems biology.

While this work was motivated by generalising methods whose aim was to scale analysis of pCTMCs, applying the GFA on a large CTMC may prove computationally prohibitive; however, approximating subsets of the state-space, or sub-systems present in the CTMC, is possible as demonstrated, which can be computationally beneficial.

On a more conceptual note, all approximations are accurate over a specific range of applicability and most only possible on certain classes of CTMCs. We therefore sought to construct a universally applicable method resting upon fundamental properties of a general CTMC.
Our method conjectures that the quality of a continuous approximation depends on how well diffusion distance can be preserved in a Euclidean space, over which the approximation evolves. Despite no significant decrease of the computational cost for CTMC analysis, we offer insight on how this property influences the quality of continuous approximations to CTMCs in general, and expect that this property may be cast into a suitable metric to quantify how continuous approximation quality varies across general CTMCs. For instance, a CTMC which is close to a pCTMC in this sense (as the ones used in the empirical section above) is expected to admit a relatively accurate continuous approximation.


Some potential paths forward become apparent under the lens of this work.
%
Firstly, our method might be optimised to accommodate particular classes of CTMCs, for example by designing specific kernels for the GP regression part. This might be an effective way to incorporate domain knowledge and further improve the quality of the geometric approximation.

Secondly, we can extend this methodology by approximating the diffusion matrix field as well as the drift vector field. This would enable us to define a diffusion process on the manifold and so construct an approximating pdf rather than a point mass. An evolving pdf will be comparable to solutions produced by Van Kampen's \emph{system size expansion}, \emph{moment closure} methods, and the chemical Langevin equation for the case of CTMCs representing chemical reaction networks.

Finally, the geometric fluid approximation produces trajectories in a low-dimensional Euclidean space, but these coordinates are not immediately interpretable as they are in the canonical fluid approximation. We have here used a Voronoi tesselation to relate continuous trajectories in the Euclidean space to those in the original discrete state-space, and used this to estimate first passage times. However, our method would benefit from further work on interpreting the dimensions of the embedding. In particular, it would be interesting to extend the work on FPTs to define methodologies to approximate more complex path properties, such as temporal logic formulae which are often encountered in computer science applications \cite{milios_probabilistic_2018,bortolussi_smoothed_2016}.






\thanks{The authors would like to thank Luca Bortolussi for the many useful discussions.
This work was supported by the EPSRC under grant EP/L027208/1.}


\bibliographystyle{RS}
\bibliography{FluidSS}

\clearpage
\appendix
\section{Diffusion maps for undirected graphs}\label{app:dm-undirected}

There exists extensive literature examining the implications of diffusion maps, as well as their limitations and strengths \cite{coifman_geometric_2005,coifman_diffusion_2006,nadler_diffusion_FP_2006,nadler_diffusion_react_2006,coifman_diffusion_stoch_2008}. What follows is therefore not an attempt to re-derive these results or convince the reader of the validity of the method, but rather to set notation and highlight the aspects that are relevant to our purposes. The exposition below is also necessary to act as a foundation for the results of Perrault-Joncas and Meilă that build upon the original concept of diffusion maps as put forth by Coifman, Lafon, Nadler, and Kevrekidis.

In \cite{nadler_diffusion_react_2006,coifman_diffusion_stoch_2008}, the authors consider a family of density-normalised (i.e.\ anisotropic) symmetric kernels
\[
k_\epsilon^{(\alpha)}(\mathbf{x}, \mathbf{y}) = \frac{k_\epsilon (\mathbf{x}, \mathbf{y})}{p^\alpha_\epsilon (\mathbf{x}) p^\alpha_\epsilon (\mathbf{y})}
\]
characterising the distance between high-dimensional points $\mathbf{x}, \mathbf{y} \in \mathcal{M} \subseteq \mathbb{R}^p$. The kernel used here is the radial basis function $k_\epsilon (\mathbf{x}, \mathbf{y}) = \exp(-d(\mathbf{x}, \mathbf{y})^2/\epsilon)$, which provides a similarity between points based on the Euclidean distance $d$ in the original space. The density-normalising factor $p^\alpha_\epsilon (\mathbf{x})$ depends on the manifold density, $p_\epsilon(\mathbf{x}) = \int k_\epsilon (\mathbf{x}, \mathbf{y} p_\epsilon(y) dy$, and the choice of the power $\alpha$ leads to transition kernels of different diffusion process operators (see below). The hyperparameter $\epsilon$ is the kernel width, which corresponds to the time elapsed between observations of a putative diffusion process (see below). For a finite set of points we can construct an adjacency matrix whose elements are given by the kernel, for a network with points as nodes and weighted undirected edges.

Assuming that the points were sampled by observing a diffusion process in the space $\mathcal{M}$, the authors then take the forward Markov transition probability kernel to be
\[
M_f^{(\alpha)}(\mathbf{x}|\mathbf{y}) = \Pr\left[\mathbf{x}(t + \epsilon) \mid \mathbf{x}(t) = \mathbf{y} \right] = \frac{k_\epsilon^{(\alpha)} (\mathbf{x}, \mathbf{y})}{d_\epsilon^{(\alpha)}(\mathbf{y})},
\]
where $d_\epsilon^{(\alpha)}(\mathbf{y}) = \int_\mathcal{M} k_\epsilon^{(\alpha)} (\mathbf{x}, \mathbf{y}) p(\mathbf{x}) d\mathbf{x}$ is the graph Laplacian normalisation factor. Since this is the transition probability for the putative continuous diffusion process evolving in the space $\mathcal{M}$, the (forward) infinitesimal diffusion operator of the process is given by
\[
\frac{\partial}{\partial t} = \mathcal{H}_f^{(\alpha)} = \lim_{\epsilon \to 0}\left[\frac{T_f^{(\alpha)} - I}{\epsilon}\right],
\]
where $I$ is the identity operator, and $T_f^{(\alpha)}$ is a (forward) transport operator defined as $T_f^{(\alpha)} [\phi] (\mathbf{x}) = \int_\mathcal{M} M_f^{(\alpha)}(\mathbf{x}|\mathbf{y}) \phi(\mathbf{y})p(\mathbf{y})d\mathbf{y}$, which evolves a function $\phi:\mathcal{M}\to\mathbb{R}$ according to $M_f^{(\alpha)}$ and the manifold measure $p(\mathbf{y}) = e^{-U(\mathbf{x})}$.

By asymptotic expansion of the relevant integrals, they show that the forward and backward operator pair is
\begin{align}
&\mathcal{H}_f^{(\alpha)} = \Delta - 2\alpha \nabla U \cdot \nabla + (2\alpha -1)(\lVert\nabla U\rVert^2 - \Delta U), &\text{and}\\
&\mathcal{H}_b^{(\alpha)} = \Delta - 2(1 - \alpha) \nabla U \cdot \nabla
\end{align}
respectively.

We then regard the adjacency matrix $W$ of a given network to be a discrete approximation of the transition kernel $k_\epsilon$ defined over continuous space. From that, we can construct discrete (in time and space) approximations to the diffusion operators $\mathcal{H}^{\alpha}$ above by performing the necessary normalisations. To retrieve the embedding coordinates for each network vertex one needs to spectrally analyse the approximation to the diffusion operator, taking the 1 to $k+1$ eigenvectors $\{\psi_j\}_{j=1}^d$ ordered by the associated eigenvalues $\{-\lambda_j\}_{j=1}^d$ with $\lambda_0=0>-\lambda_1\geq-\lambda_2\geq\cdots \geq-\lambda_d$, to be the vertices' coordinates in the first $k < d$ dimensions of the embedding. The first eigenvector is discarded as a trivial dimension where every vertex has the same coordinate by construction. Thus, the $k$-dimensional diffusion map at time $t$ is defined as:
\[
\Psi_k^t(\mathbf{x}) := \left(e^{-\lambda_1 t}\psi_1(\mathbf{x}), e^{-\lambda_2 t}\psi_2(\mathbf{x}), \dots, e^{-\lambda_k t}\psi_k(\mathbf{x})\right),
\]
where we have discarded $\psi_0$ associated with $\lambda_0 = 0$ as a trivial dimension. The time parameter $t$ refers to the diffusion distance after time $t$ which is preserved as Euclidean distance in the embedding space. Trivially, as $t\to\infty$ all network nodes are mapped to the same point since the diffusion distance vanishes.

The parameter $\alpha$ adjusts the effect that the manifold density has on the diffusion process. Choosing $\alpha=1$ recovers the Laplace-Beltrami operator $\Delta$ as the backward diffusion operator, if the points approximately lie on a manifold $\mathcal{M}\subset \mathbb{R}^d$. Thus, the diffusion map corresponds to an embedding of the points unaffected by the manifold density (such that if two different networks were sampled from the same manifold $\mathcal{M}$ but with different densities, we would recover consistent positions of the points on $\mathcal{M}$). Choosing $\alpha=0$ is equivalent to the \emph{Laplacian eigenmaps} method which preceded diffusion maps \cite{belkin_laplacian_2003}. If the vertices are sampled uniformly from the hidden manifold, Laplacian eigenmap becomes equivalent to analysing the Laplace-Beltrami operator, and so constructing a diffusion map with $\alpha=1$ and with $\alpha=0$ will recover the same embedding \cite{coifman_diffusion_2006}.

Consider now an Itô stochastic differential equation (SDE) of the form
\begin{equation}\label{eq:SDE}
\dot{\mathbf{x}} = \bm{\mu}(\mathbf{x}) + \bm{\sigma} \dot{\mathbf{w}},
\end{equation}
where $\mathbf{w}_t$ is the $d$-dimensional Brownian motion. A probability distribution over the state-space of this system $\phi(\mathbf{x}, t)$ with condition $\phi(\mathbf{x}, 0) = \phi_0(\mathbf{x})$, evolves forward in time according to the Fokker-Planck equation (FPE), also known as the Kolmogorov forward equation (KFE):
\begin{align}
\partial_t \phi(\mathbf{x}, t) = - \sum_i \partial_i \left[\mu_i(\mathbf{x}) \phi(\mathbf{x}, t)\right] + \sum_i \sum_j \partial_i \partial_j \left[\frac{1}{2} \sigma_i \sigma_j \phi(\mathbf{x}, t)\right],
\end{align}
with the sums running over all $d$ dimensions and $\partial_i$ denoting partial derivatives with respect to the $i$th dimension ($\partial_i = \partial / \partial x_i$) \cite{gardiner_stochastic_2009}. Similarly, the probability distribution $\psi(\mathbf{y}, s)$ for $s \leq t$ and condition $\psi(\mathbf{y}, t) = \psi_t(\mathbf{x})$ satisfies
\begin{align}
- \partial_s \psi= \bm{\mu} \cdot \nabla \psi + \frac{1}{2} \bm{\sigma} \bm{\sigma}^\top\Delta \psi,
\end{align}
where the differentiations are with respect to $\mathbf{y}$. Terms in the backward FPE become directly identifiable with the backward operator $\mathcal{H}_b^{(\alpha)}$ if we take $\bm{\sigma} = \sqrt{2}\mathbf{I}$ and $\bm{\mu} = 2(1 - \alpha) \nabla U$.

The original formulation of diffusion maps, as described above, assumes a symmetric kernel $k_\epsilon(\mathbf{x}, \mathbf{y}) = k_\epsilon(\mathbf{y}, \mathbf{x})$. Given a CTMC with a symmetric generator matrix $Q$, the methodology laid out so far would be sufficient to recover an embedding for the states on a continuous compact manifold $\mathcal{M}$, on which we can define an SDE approximation to the Markov jump process of the CTMC. Encouragingly, it has also been shown that the jump process would satisfy the reflecting (no flux) conditions on the manifold boundary $\partial\mathcal{M}$, as required by a diffusion FP operator defined on such a manifold --- i.e.\ for a point $\mathbf{x} \in \partial\mathcal{M}$ where $\mathbf{n}$ is a normal unit vector at $\mathbf{x}$, and a function $\psi:\mathcal{M} \to \mathbb{R}$,
\[
\left. \frac{\partial\psi(\mathbf{x})}{\partial \mathbf{n}}\right|_{\partial\mathcal{M}} = 0.
\]

\section{Embedding unweighted, undirected, grid graphs}
\label{app:grids}
Taking the case of a pCTMC produced by a particular class of chemical reaction networks, we show that the embedding produced by \emph{Laplacian eigenmaps} \cite{belkin_laplacian_2003} (equivalent to diffusion maps with $\alpha=0$) for the unweighted, undirected transition matrix, is consistent in some respect to the canonical (manual) embedding for the fluid limit of chemical reaction systems. This implies that we ignore any density information of the vertices (states) on the manifold, and any directional component. We will later return to how this information affects our results.

\paragraph{Laplacian eigenmaps embedding}
Assume that we have symmetric similarity matrix $W$ between $n$ points. We construct the Laplacian matrix $L = D - W$, with $D_{ii} = \sum_j W_{ji}$.
The Laplacian eigenmaps algorithm solves the minimisation problem
\begin{align}
&\argmin_{\Upsilon^\top D \Upsilon = I}
\frac{1}{2} \sum_{i, j} \lVert\mathbf{y}^{(i)} - \mathbf{y}^{(j)}\rVert_2^2 ~ W_{ij} \\
&= \argmin_{\Upsilon^\top D \Upsilon = I} \mathrm{Tr} (\Upsilon^\top L \Upsilon),
\end{align}
where $\mathbf{y}^{(i)}$ is the $i$th row of $\Upsilon$, and the constraint $\Upsilon^\top D \Upsilon = I$ serves to exclude the trivial solution of mapping everything to the same point. The solution $\Upsilon \in \mathbb{R}^{n \times m}$ is a matrix with each column vector corresponding to the $m$-dimensional coordinate embedding of each datum ($m<n$). It is shown that the solution to the problem is the eigenvector matrix corresponding to the $m$ lowest eigenvalues of $L\mathbf{y} = \lambda D \mathbf{y}$, excluding the $\lambda=0$ solution.

This emphasis on preserving local information allows us to appropriate the algorithm for embedding the network of states without having to calculate global state separation --- i.e.\ by using only neighbouring state similarities as represented in $Q$. For a CTMC described by a transition matrix $Q$, we transform $Q$ to be an adjacency matrix between the nodes (states) of the network (CTMC) by placing an undirected edge of weight 1 between states which are separated by a single transition and 0 otherwise:
\begin{align}
W_{ij} = 1 - \delta_{0, Q_{ij}} \delta_{0, Q_{ji}}.
\end{align}

If the network is connected and $m$ (the dimensionality for the embedding space) is picked appropriately, the algorithm will attempt to preserve local dimensions and therefore global ones if the network fits in that $m$ space. If $m$ is chosen higher than necessary, some states which are far apart might be placed closer together in the embedding, but local distances will still be preserved.

\paragraph{The unweighted Laplacian fluid approximation}

The proof for Theorem~\ref{gfa:thm:consistency} is laid out here. It involves the construction of an undirected, unweighted graph with adjacency matrix $W$ from the $Q$ matrix of a specific kind of pCTMCs, as shown above. Explicit eigenvectors of the Laplacian $L$ of this graph give analytic coordinates for the vertices of $Q$ in some space $\mathbb{R}^d$. A drift vector field is inferred on this space using Gaussian process regression, from $Q$ and the embedding coordinates. We show from these how conditions for a fluid approximation are met, as stated in Section~\ref{gfa:sec:theory:established}. Specifically, we show how \emph{initial conditions converge}, \emph{mean dynamics converge}, and \emph{noise converges to zero} (via Taylor expansion of the relevant analytic coordinates), in the same way as in the canonical embedding of such a pCTMC resulting from \emph{hydrodynamic scaling}.

\paragraph{Theorem~\ref{gfa:thm:consistency}}{\it Let $\mathcal{C}$ be a pCTMC, whose underlying transition graph maps to a multi-dimensional grid graph in Euclidean space using the canonical \emph{hydrodynamic scaling} embedding. The \emph{unweighted Laplacian fluid approximation} of $\mathcal{C}$ coincides with the canonical fluid approximation in the hydrodynamic scaling limit.}

\begin{proof}
We examine a particular case of pCTMCs, produced by allowing reactions that only change the count of a single species per reaction. This produces an adjacency matrix $W$ for the network of states describing a grid network in $d$ dimensions. Following the derivation for the eigenvectors of the Laplacian $L$ of such a network presented in \cite{klopotek_spectral_2017}, we find that the lowest eigenvalue $\lambda_1$ (excluding $\lambda_0=0$) is degenerate ($\lambda_1 = \lambda_{\{2,\dots, d\}}$), and associated with $d$ eigenvectors $\mathbf{v}_{j},~j\in\{1,\dots, d\}$. Their elements are
\begin{align}
\mathbf{v}_{j,[x_1,\dots, x_d]} =
\cos\left(\frac{\pi }{n_j}\left(x_j - \frac{1}{2}\right)\right)
\end{align}
where the index $[x_1,\dots, x_d]$ is the mapping of the node to its integer grid coordinates. Therefore, the embedded $j$th coordinate of a node is $\cos(\pi/n_j(x_j - 1/2))$, where $x_j \in \{1, \dots, n_j\}$ is the integer grid position of the node in that $j$ dimension. We observe that away from the boundaries (i.e.\ near the centre of the grid $x\approx n/2$) and for large $n$, the argument of $\cos$ is near $\pi/2$, so we approach the linear part of $\cos$. This means that near the centre states are almost uniformly distributed, as in the canonical embedding.

We define the volume $\Omega_U([x_1,\dots, x_d])$ for a state with grid coordinates $[x_1,\dots, x_d]$ in the network, to be the volume of the polygon ($n$-orthotope) whose vertices are that state and the next state along each grid dimension:
\begin{align}
\Omega_U([x_1,\dots, x_d]) &= \prod_j\left( \mathbf{v}_{j,[x_1,\dots,x_j +1,\dots, x_d]} - \mathbf{v}_{j,[x_1,\dots,x_j ,\dots, x_d]} \right)  \\
    &= \prod_j \left[ \cos\left(\frac{\pi }{2n_j}\left(2x_j + 1\right)\right)
        - \cos\left(\frac{\pi }{2n_j}\left(2x_j - 1\right)\right) \right].
\end{align}
We then observe that $\lim_{n \to \infty} \Omega_U = 0$ for all states; this satisfies the convergence condition of initial states for a fluid approximation.

We define dynamics by means of a drift field $\expval{b}:U\to\mathbb{R}^d$. The function is inferred using Gaussian process regression, $b(\cdot) \mid Q \sim \mathcal{GP}(m(\cdot)\mid Q, k(\cdot, \cdot)\mid Q)$, such that it is a Lipschitz field. This satisfies the convergence condition of mean dynamics for a fluid approximation. In the canonical embedding of a pCTMC, the drift vector field is a polynomial function $f_p \in L^2(U)$ over the concentration space. Away from the boundaries, the Laplacian embedding approaches this canonical embedding. As $n\to\infty$, the inferred field in this region will tend to the same polynomial function:
\[
\expval{b} \to f_p \quad,
\]
as the Gaussian process can approximate any function in $L^2(U)$ arbitrarily well.

Finally, the conditions for noise converging to zero are trivially met, since embedding distances $\gamma$ are at most $\order{n^{-1}}$:
\begin{align*}
\gamma &= \cos\left(\frac{\pi }{2n_j}\left(2x_j + 1\right)\right)
        - \cos\left(\frac{\pi }{2n_j}\left(2x_j - 1\right)\right)\\
    &= 1 - \frac{1}{2!}\left(\frac{\pi }{2n_j}\left(2x_j + 1\right)\right)^2 + \frac{1}{4!}\left(\frac{\pi }{2n_j}\left(2x_j + 1\right)\right)^4 - \dots \\
    &\quad - 1 + \frac{1}{2!}\left(\frac{\pi }{2n_j}\left(2x_j - 1\right)\right)^2 - \frac{1}{4!}\left(\frac{\pi }{2n_j}\left(2x_j - 1\right)\right)^4 + \dots \\
    &= \order{n_j^{-1}},
\end{align*}
and $n = \sum_j n_j$, such that $\gamma^2 = \order{n^{-2}}$.

Thus the criteria for \emph{fluid approximation} of this pCTMC are satisfied. Further, for some region of the state-space and in the limit of infinite states, this construction is consistent with the embedding and dynamics recovered by \emph{hydrodynamic scaling}, the canonical \emph{fluid approximation} of a pCTMC. This concludes our proof.
\end{proof}

\section{Diffusion maps for directed graphs}\label{app:dm-directed}

Our focus necessarily shifts on embedding an arbitrary CTMC with no symmetry condition on $Q$. Following Perrault-Joncas and Meilă \cite{perrault-joncas_directed_2011} assume that we observe a graph $G$, with nodes sampled from a diffusion process on a manifold $\mathcal{M}$ with density $p=e^{-U}$ and edge weights given by the (non-symmetric) kernel $k_\epsilon$. The directional component of the kernel is further assumed to be derived from a vector field $\mathbf{r}$ on $\mathcal{M}$ without loss of kernel generality. As the authors saliently put it: ``The question is then as follows: can the generative process' geometry $\mathcal{M}$, distribution $p=e^{-U}$, and directionality $\mathbf{r}$, be recovered from $G$?''

In the same manner as for the original formulation of diffusion maps a set of backward evolution operators are derived, the two relevant ones being:
\begin{align}
&-\partial_t = \mathcal{H}_{aa}^{(\alpha)} = \Delta + \left(\mathbf{r} - 2(1 - \alpha)\nabla U \right)\cdot \nabla, &\text{and} \\
&-\partial_t = \mathcal{H}_{ss}^{(\alpha)} = \Delta - 2(1 - \alpha)\nabla U \cdot \nabla.
\end{align}
To construct this family of operators, the kernel is first decomposed into its symmetric $h_\epsilon$ and anti-symmetric $a_\epsilon$ parts,
\[
k_\epsilon^{(\alpha)}(\mathbf{x}, \mathbf{y}) = \frac{k_\epsilon(\mathbf{x}, \mathbf{y})}{p^{\alpha}_\epsilon(\mathbf{x}) p^{\alpha}_\epsilon(\mathbf{y})} = \frac{1}{p^{\alpha}_\epsilon(\mathbf{x}) p^{\alpha}_\epsilon(\mathbf{y})}\left[ h_\epsilon(\mathbf{x}, \mathbf{y}) + a_\epsilon(\mathbf{x}, \mathbf{y}) \right],
\]
and further normalised according to either the asymmetric $d^{(\alpha)}_\epsilon(\mathbf{x}) = \int_\mathcal{M} k_\epsilon^{(\alpha)}(\mathbf{x}, \mathbf{y}) p(\mathbf{y}) d\mathbf{y}$, or symmetric outdegree distribution $\tilde{d}^{(\alpha)}_\epsilon(\mathbf{x}) = \int_\mathcal{M} h_\epsilon^{(\alpha)}(\mathbf{x}, \mathbf{y}) p(\mathbf{y}) d\mathbf{y}$.
The subscript indices denote the type of kernel used to construct the operator and the outdegree distribution used to normalise it (such that $\mathcal{H}_{aa}$ associates to the full asymmetric kernel $k_\epsilon^{(\alpha)}$ normalised with asymmetric degree distribution $p_\epsilon$, and so on).

Discrete approximations for these operators can be constructed for an asymmetric kernel matrix of distances between $N$ high-dimensional points, $\mathbf{W} \in \mathbb{R}^{N \times N}$. The symmetric matrix $H_{ss}^{(1)} \in \mathbb{R}^{N \times N} $ can be extracted and the necessary eigen-decomposition carried out to yield an embedding, where $\lim_{N\to\infty} H_{ss}^{(1)} = \mathcal{H}_{ss}^{(1)} = \Delta$.
However, given the infinitesimal generator of a CTMC $Q$, we do not have access to $\mathbf{W}$, but rather to the discrete approximation of the final evolution operator, $\lim_{N\to \infty} Q = \mathcal{H}_{aa}^{(\alpha)}$. In order to recover the initial kernel matrix $\mathbf{W}$ that gave rise to $Q$, we take $\alpha=0$, a uniform measure on the manifold $U(\mathbf{x})=0 \implies p(\mathbf{x})=1$, and a small value for $\epsilon$. This makes the transformations from $\mathbf{W}_\epsilon$ to $Q$ reversible, since
\begin{align}
&Q = \lim_{\epsilon\to0} \left[\frac{T^{(\alpha=0)}_\epsilon - \mathbf{I}}{\epsilon}\right], \text{ and}\\
&T^{(\alpha=0)}_\epsilon = D^{-1} \mathbf{W}_\epsilon, \text{ such that}\\
&\mathbf{W}_\epsilon = D (\mathbf{I} + \epsilon Q.)
\end{align}
In the above, $D$ is a diagonal matrix which forces the diagonal of $\mathbf{W}_\epsilon$ to be 1, as expected from a distance-based kernel matrix.
The final step is the familiar \emph{uniformisation} procedure which approximates a CTMC with a DTMC. The choice of $\epsilon < \left(\max_i |Q_{ii}|\right)^{-1}$ determines the quality of approximation (the smaller the better).

Once the kernel matrix $\mathbf{W}_\epsilon$ is recovered we can proceed to construct the operators $\Delta = \mathcal{H}_{ss}^{(1)}$ and $\left( \mathcal{H}_{aa}^{(0)} - \mathcal{H}_{ss}^{(1)} \right) = \left(\mathbf{r} - 2\nabla U \right)\cdot \nabla$, which are used to embed the state-space on a manifold $\mathcal{M} \in \mathbb{R}^d$, and endow it with the advective field $\bm{\mu} = \left(2\nabla U + \mathbf{r}\right)$ in the Kolmogorov backward equation, respectively. See Algorithm~\ref{alg:gfa} for procedural details of the \emph{geometric fluid approximation}.

\end{document}